\documentclass{aastex631}

\graphicspath{{./}{figures/}}

\shorttitle{Sun-as-a-star Analyses of Various Solar Active Events}
\shortauthors{Otsu et al.}

\begin{document}

\title{Sun-as-a-star Analyses of Various Solar Active Events Using H$\alpha$ Spectral Images Taken by SMART/SDDI }

\author{Takato Otsu}
\affiliation{Department of Astronomy, Kyoto University, Sakyo, Kyoto, Japan}

\author{Ayumi Asai}
\affiliation{Astronomical Observatory, Kyoto University, Sakyo, Kyoto, Japan}

\author{Kiyoshi Ichimoto}
\affiliation{Astronomical Observatory, Kyoto University, Sakyo, Kyoto, Japan}

\author{Takako T. Ishii}
\affiliation{Astronomical Observatory, Kyoto University, Sakyo, Kyoto, Japan}

\author{Kosuke Namekata}
\affiliation{ALMA Project, NAOJ, NINS, Osawa, Mitaka, Tokyo, Japan}

\correspondingauthor{Takato OTSU}
\email{t.otsu@kusastro.kyoto-u.ac.jp}
\begin{abstract}
Sun-as-a-star analyses, in which observational data is spatially integrated, are useful for interpreting stellar data. For future applications to stellar observations, we performed Sun-as-a-star analyses of H$\alpha$ spectra for various active events on the Sun, not only for flares and filament eruptions/surges on the solar disk, but also for eruptions of off limb prominences using H$\alpha$ spectral images taken by the Solar Magnetic Activity Research Telescope / Solar Dynamics Doppler Imager (SMART/SDDI) at Hida Observatory, Kyoto University. All the analyzed events show emission relative to the pre-event state and the changes in their H$\alpha$ equivalent widths are all on the orders of 10$^{-4}$ {\AA}. Sun-as-a-star H$\alpha$ spectra exhibit different features depending on the causes of the emission: (i) Flares show emission at the H$\alpha$ line center, together with red asymmetry and line broadening, as reported in a previous study. (ii) Filament eruptions with and without flares show emission near the H$\alpha$ line center, accompanied by blue-/red-shifted absorption. Notably, disappearance of dark filaments leads to the apparent enhancement of the H$\alpha$ line center emission. (iii)  Eruptions of off limb prominences show blue-/red-shifted emission. 
These spectral features enable us to identify the active phenomena on Sun-like stars. 
We have also found that even the filament eruptions showing red-shifted absorptions in Sun-as-a-star H$\alpha$ spectra lead to coronal mass ejections (CMEs).
This result suggests that even if the falling components of stellar filament eruptions are detected as red-shifted absorptions in H$\alpha$ spectra, such stellar filament eruptions may also develop into CMEs.


\end{abstract}

\keywords{sun, flare, filament eruption, prominence eruption, CME}

\section{Introduction \label{s:1}}
Solar flares are explosive brightenings in the solar atmosphere. They are thought to be caused by the
conversion of magnetic energy into kinetic, thermal, and non-thermal energies via magnetic reconnection \citep{ShibataMagara2011}. In addition, filament/prominence eruptions and coronal mass ejections (CMEs) sometimes acommpany solar flares.
The interaction between CMEs and Earth's magnetosphere could produce geomagnetic storms, which have serious impacts 
on the modern human life \citep[e.g.,][]{AirapetianETAL2020, CliverETAL2022}.

Like the Sun, various stars also show sudden brightenings called stellar
flares. In recent years, superflares---which are stellar flares emitting more than $10^{33}$ ergs---have been observed even in stars similar to the Sun \citep{MaeharaETAL2012,ShibayamaETAL2013,TuETAL2020,OkamotoETAL2021}. In order
to understand flares including superflares in unified way, stellar flares are being investigated actively \citep[e.g.,][]{KowalskiETAL2013,TsuboiETAL2016,HondaETAL2018,NotsuETAL2019,VidaETAL2019,MuhekiETAL2020,NamekataETAL2020b,MaeharaETAL2021,NamekataETAL2022a,NamekataETAL2022b}. 
Stellar superflares may be accompanied by much larger CMEs than those of the Sun, which are thought to have severe impacts
on the exoplanets around the host stars \citep[e.g.,][]{AirapetianETAL2020}. Consequently, superflares and its associated CMEs have been gathering 
an attention, not only from physical viewpont but also from the perspective of the habitability of exoplanets.

The surface of the Sun can be observed with high spatial resolution, while the surfaces of distant stars cannot be resolved spatially. In recent years, the spatially resolved data of the Sun have been utilized to aid the analysis of stellar data \citep[e.g.,][]{HarraETAL2016, ToriumiETAL2020,VeronigETAL2021,NamekataETAL2022a,NamekataETAL2022b,ToriumiAirapetian2022, XuETAL2022}. For comparison with the stellar data, solar data are integrated spatially, and such an analysis is called “Sun-as-a-star analysis”. \citet{NamekataETAL2022a} performed Sun-as-a-star analyses using H$\alpha$ images of two solar flares accompanied by filament eruption/surge to interpret the H$\alpha$ spectra of a stellar flare. Since both solar and stellar spectra have similar characteristics, the authors concluded that a stellar filament eruption associated with a superflare had been detected. In addition, \citet{NamekataETAL2022c} analyzed the Sun-as-a-star H$\alpha$ spectrum of an M-class solar flare that exhibited dominant emission from flare ribbons, and they showed that red asymmetry and line broadening can be seen even in a Sun-as-a-star spectrum, providing evidence of chromospheric condensation. This is how Sun-as-a-star analyses of H$\alpha$ spectra are useful for investigating the dynamics of flares and plasmas on a stellar surface. Also, in contrast to extreme ultraviolet (EUV) radiation that are also useful to study stellar activities, especially stellar CMEs \citep{HarraETAL2016, XuETAL2022} but cannot reach Earth from distant stars due to interstellar absorption \citep{RumphETAL1994}, we can observe H$\alpha$ radiation even coming from distant stars. 
It is therefore important to investigate H$\alpha$ spectra by performing Sun-as-a-star analyses not only for flares and filament eruptions/surges but also for various other active phenomena on the Sun.

So far, Sun-as-a-star analyses of H$\alpha$ spectra have been performed for only three solar active events \citep[flares and filament eruptions/surges;][]{NamekataETAL2022a,NamekataETAL2022c}. 
In this paper, we report the results of Sun-as-a-star analyses of H$\alpha$ spectra for more various active events on the Sun. In addition to flares and filament eruptions/surges, there is a wide variety of chromospheric plasma phenomena---such as out-of-limb prominence eruptions \citep[e.g.,][]{Parenti2014} and post-flare loops \citep[e.g.,][]{JingETAL2016}---and we have included these phenomena, which have not been analyzed previously from the Sun-as-a-star viewpoint. Our main goal is to clarify the correspondence between various active events and the characteristics of the H$\alpha$ spectra in Sun-as-a-star analyses, for future applications to stellar observations. We have mainly followed the method of Sun-as-a-star analyses described by \citet{NamekataETAL2022a}, while we have incorporated our own calibration method in this paper. We introduce the observational data in Section \ref{s:Ob} and explain the analytical method in Section \ref{s:Me}. In Section \ref{s:Re}, we report the results and discuss them. Finally, we present a summary and implications to stellar observations in Section \ref{s:SD}. 

\section{Observations}\label{s:Ob}
The Solar Dynamics Doppler Imager \citep[SDDI;][]{IchimotoETAL2017} is installed on the Solar Magnetic Activity Research Telescope \citep[SMART;][]{UenoETAL2004} at Hida Observatory, Kyoto University . It takes full-disk solar images at 73 wavelengths from the H$\alpha$ line center $-9.0$ {\AA} to the H$\alpha$ line center $+9.0$ {\AA} with a time cadence of 12$-$16 sec and a pixel size of 1.23 arcsec. The SDDI has been observing the Sun routinely since 2016 May and has stored the H$\alpha$ spectral images of various active events. 

We have selected relatively prominent events for study in this paper: flares with \textit{GOES} (Geostationary Operational Environmental Satellites) X-ray classes of $\geq$ C8.0 and plasma eruptions with large spatial scales (on the order of approximately 100 Mm). We excluded some events because of clouds or terribly bad seeing. We ultimately selected nine events. For these events, we summarize in Table\ref{table1} the dates of the observations, the \textit{GOES} peak time, \textit{GOES} class, location, and event features. 
We also summarize in Table \ref{table1} the occurrence of CMEs based on the data of the Large Angle and Spectrometric Coronagraph \citep[LASCO;][]{BruecknerETAL1995} on board the Solar and Heliospheric Observatory (SOHO) and the catalogue of filament disappearance in \citet{SekiETAL2019b}.
In addition to the SDDI data, for some events we used data taken with the Atmospheric Imaging Assembly \citep[AIA;][]{LemenETAL2012AIA} on board the Solar Dynamics Observatory \citep[SDO;][]{PesnellETAL2012} to ensure the event scenarios. In the following, we provide observational classifications of the events studied in this paper. Note that this classification is not essential but is just a convenience for explanation. The details of each event are given in Section \ref {s:Re}.

Events (1)$-$(4) were associated with flares ($\geq$ C8.0). Event (1) is the same as that analyzed by \citet{NamekataETAL2022a}. Although this event has already been analyzed in detail, it is included in this paper again with the intent of comparing it to other events. 
Events (5)$-$(7) were filament eruptions associated with weak flares (B-class). In these three events, dark filaments were observed on the solar disk in the H$\alpha$ line center images, and they erupted with partial drainage.
Events (8) and (9) were prominence eruptions to the outside of the solar limb. The footpoints of the flares from these events were occulted, and their \textit{GOES} data are therefore not summarized in Table \ref{table1}. 

\begin{deluxetable*}{ccccccc}
\label{table1}
\tablenum{1}
\tablecaption{List of events analyzed in this paper}
\tablewidth{0pt}
\tablehead{
\colhead{Event} &
\colhead{Date (UT)} & \colhead{\textit{GOES} Peak Time (UT)}&\colhead{\textit{GOES} Class} & \colhead{Location}&\colhead{Event features}&
\colhead{CME}
}
\decimals
\startdata
(1) & 2017 Apr. 2 & 02:46:00 & C 8.0 & S12 W08 & flare + surge & no ?\\
(2) & 2017 Apr. 2 & 08:02:00 & M 5.3 & N12 W59  & flare + eruption & yes\\
(3) & 2017 Sept. 8 & 07:49:00 & M 8.1 & S09 W70 & flare (two-step) + eruption & no ?\\
(4) & 2021 Apr. 19 -- 20 & Apr. 19 23:42:00 & M 1.1 & 
S24 E14 & flare + coronal rain ? & yes\\
(5) & 2016 Nov. 5 & 04:52:00 & B 1.1 & N08 W32 &two-ribbon flare + filament eruption & yes\\
(6) & 2017 Feb. 19 & 05:41:00 & B 3.0 & N11 E15 &two-ribbon flare + filament eruption & yes\\
(7) & 2017 Apr. 23 & 05:55:00 & B 1.8 & N13 E33 &two-ribbon flare + filament eruption & yes\\
(8) & 2017 June 19 &         &            & NE limb & prominence eruption & yes\\
(9) & 2021 May 5 & & & NE limb &prominence eruption (two-step) & yes\\
\enddata
\tablecomments{The flaring footpoints of event (8) and event (9) were occulted so we did not include the \textit{GOES} class and Peak Time of these two events in Table \ref{table1}.
The occurrence of CMEs for event (1)$-$(4), event (8), and event (9) is based on SOHO/LASCO data , and it for event (5)$-$(7) is referred from \citet{SekiETAL2019b} . 
The term “no ?” means that we cannot decide the occurrence of CMEs from SOHO/LASCO data.
Note that although event (4) was not associated with any H$\alpha$ eruptions, it would be followed by a hotter eruption. }
\end{deluxetable*}

\section{Method \label{s:Me}}
Because the brightening of a solar flare is much smaller than the radiation from the full solar disk, it is difficult to identify clear changes in Sun-as-a-star spectra that are simply integrated over the full disk. We therefore integrated the H$\alpha$ spectra over partial target regions (TRs) which that contain the active events we consider. This is equivalent to assuming that there are no other variations except for that in the TR. We converted these spatially integrated spectra in TRs to “virtual”  Sun-as-a-star spectra \citep[see][]{NamekataETAL2022a} by normalizing them with the full-disk integrated solar continuum. The details of this method are follows.

First, we define $f(t,\lambda,A)$ as an H$\alpha$ spectrum that is simply integrated over a region denoted by A:

\begin{equation}
f(t,\lambda,A)=\int_{\mathrm{A}}{I(t,\lambda,x,y)}\mathrm{d}x\mathrm{d}y,
\end{equation}

where $I(t,\lambda,x,y)$ is the intensity as a function of the time of the observation $t$, the wavelength $\lambda$, and the spatial position $(x,y)$. In order to suppress intensity fluctuations coming from temperature variations of the SDDI, we normalized $f(t,\lambda,A)$ by its continuum level at each time:

\begin{equation}
\label{F}
F(t,\lambda,A)=\frac{f(t,\lambda,A)}{f(t,\lambda_{cont.},A)}\times f(t_0,\lambda_{cont.},A),
\end{equation}

where $t_0$ is a pre-event time and $\lambda_{cont}$ is a continuum wavelength near the H$\alpha$ line. The next step is normalization by a quiet region (QR). We used Eq. (\ref{F}) to obtain $F(t,\lambda,A=TR)$ and $F(t,\lambda,A=QR)$. We then
calculated the H$\alpha$ spectrum $F_{TR}(t,\lambda)$ integrated over the TR and normalized by the QR data: 

\begin{equation}
\label{F_TR}
F_{TR}(t,\lambda)=\frac{F(t,\lambda,A=TR)}{F(t,\lambda,A=QR)}\times F(t_{0},\lambda,A=QR).
\end{equation}
This normalization suppresses influences coming from the observational environment, such as from changes in solar altitudes or from the Earth's atmospheric fluctuations. After the two normalizations given by Eq.(\ref{F}) and Eq.(\ref{F_TR}), we subtracted the pre-event data from $F_{TR}(t,\lambda)$ to obtain the change due to the active events:

\begin{equation}
\Delta F_{TR}(t,\lambda)=F_{TR}(t,\lambda)-F_{TR}(t_{0},\lambda),
\end{equation}

Finally, we normalized $\Delta F_{TR}(t,\lambda)$ by the full-disk solar continuum $F(t_{0},\lambda_{cont.},A=\mathrm{full~disk})$:

\begin{equation}
\label{dS}
\Delta S(t,\lambda)=\Delta F_{TR}(t,\lambda)/F(t_{0},\lambda_{cont.},A=\mathrm{full~disk}),
\end{equation}

The resulting normalized pre-event-subtracted H$\alpha$ spectrum $\Delta S(t,\lambda)$ represents the ratio of the spectral changes coming from active events to the solar irradiance (full-disk continuum). If there is no other variation except for the target event, we can calculate $\Delta S(t,\lambda)$ by using the observational data as if we were observing the Sun as a distant star. Therefore, $\Delta S(t,\lambda)$ can be regarded as a Sun-as-a-star H$\alpha$ spectrum. 
We also calculated the differenced equivalent width: $\Delta \mathrm{EW}_{\mathrm{H}\alpha\pm\Delta \lambda}=\int_{\text{H}\alpha-\Delta\lambda}^{\text{H}\alpha+\Delta\lambda}{\Delta S(t,\lambda)\mathrm~{d}\lambda}$ where $\mathrm{H}\alpha\pm\Delta\lambda$ means the wavelength of H$\alpha$ line center $\pm$ $\Delta\lambda$ [\AA]. The differenced equivalent width corresponds to the total change in the H$\alpha$ spectrum.

\section{Results \& Discussion \label{s:Re}}
Here we introduce the time evolution of each of the nine events in spatially resolved images as well as the Sun-as-a-star H$\alpha$ spectra $\Delta S(t,\lambda)$ of the nine events illustrated as dynamic spectra; i.e., as color maps of $\Delta S(t,\lambda)$. We also show the differenced equivalent widths $\Delta \mathrm{EW}_{\mathrm{H}\alpha\pm\Delta \lambda}$ and the \textit{GOES} soft X-ray (SXR) light curves.
We discuss the causes of the H$\alpha$ spectral formations from the Sun-as-a-star viewpoint by referring to the spatially resolved images.
We use the line centers calculated by fitting quadratic functions to the line cores of quiet-region data as the H$\alpha$ line center 6562.8 {\AA} in the dynamic spectra. The absolute wavelengths of the H$\alpha$ spectra observed by SMART/SDDI are not accurate because the Lyot filter, which is the main component of SMART/SDDI, experiences temperature drift. Therefore, corrections to the wavelength zero point are necessary in order to obtain accurate Doppler velocities. 

\subsection{Event (1), a C8.0 flare on 2017 April 2}
\begin{figure}[htbp]
\centering
\includegraphics[width=14cm]
{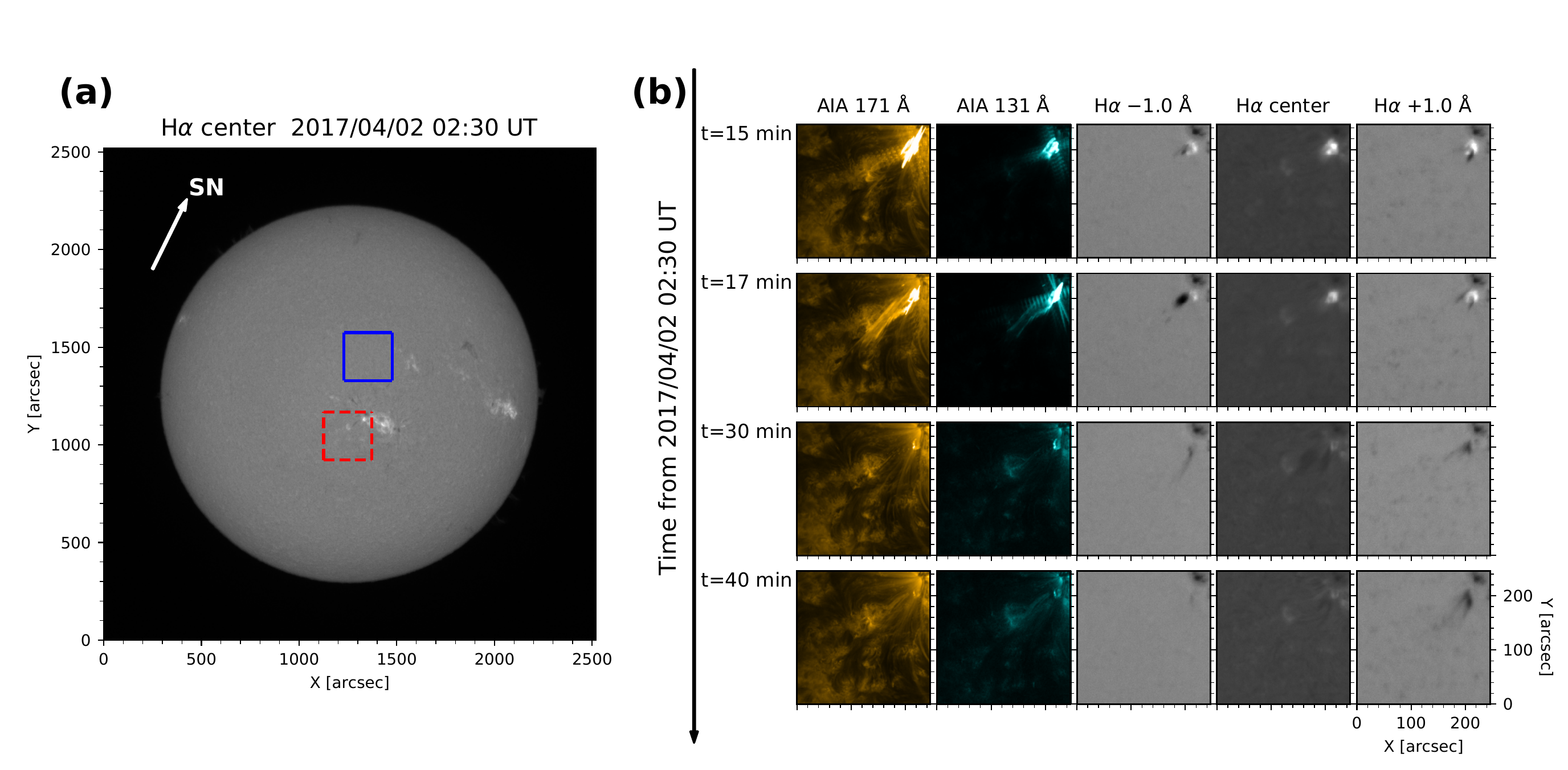}
\caption{Overview of event (1), which occurred on 2017 April 2. (a) A full-disk solar image at the H$\alpha$-center wavelength observed by SMART/SDDI. Celestial north is up, and west is to the right. The red dashed region is a target region (TR), and the blue solid region is a quiet region (QR). The direction of solar north is shown by the white arrow labeled “SN”. (b) The time evolution of the C8.0 flare on 2017 April 2. The field of view in each panel corresponds to the TR in Figure \ref{IM_C80} (a). In each row,
the images at the times t=15 min, t=17 min, t=30 min and t=40 min measured from 02:30 UT on 2017 April 2 are shown from top to bottom. In each column, the images AIA 171~{\AA}, AIA 131~{\AA}, H$\alpha-1.0$~{\AA}, H$\alpha$ center and H$\alpha+1.0$~ {\AA} are shown from left to right.}
\label{IM_C80}
\end{figure}

Figure \ref{IM_C80} shows an overview of event (1). In this event, a C8.0 flare occurred on 2017 April 02:30 UT in NOAA 12645 near the disk center [the red dashed region in Figure \ref{IM_C80} (a)]. This flare was followed by an H$\alpha$ surge [Figure \ref{IM_C80} (b)].

Figure \ref{fig:2017_0402_C} shows the result of the Sun-as-a-star analysis of the H$\alpha$ spectra of this C8.0 flare. In the dynamic spectrum [Figure \ref{fig:2017_0402_C} (a)], the emission appears near the H$\alpha$ line center. In the early phase of this emission, it shows line broadening. A strongly shifted absorption---transforming from a blue shift to a red shift---follows this emission. 
This strongly shifted absorption in the dynamic spectrum corresponds to the plasma eruption and the downflow on the solar disk, i.e., the surge. 
The $\Delta \mathrm{EW}_{\mathrm{H}\alpha\pm3.0\text{\AA}}$ also shows strong absorption (on the order of $\times10^{-4}$ {\AA}), and its time evolution deviates from the \textit{GOES} SXR light curve [Figure \ref{fig:2017_0402_C} (b)]. The flaring amplitude is $\sim3\times10^{-4}$ {\AA}, and the amplitude of the absorption is $\sim2\times10^{-4}$ {\AA}.

These results are essentially consistent with the same analysis of the same event by \citeauthor{NamekataETAL2022a} (\citeyear{NamekataETAL2022a}; see their Supplementary Information for a more detailed analysis of this event.). This consistency indicates the robustness of the Sun-as-a-star analysis method.

\subsection{Event (2), an M5.3 flare on 2017 April 2}
\begin{figure}
\centering
\includegraphics[width=14cm]
{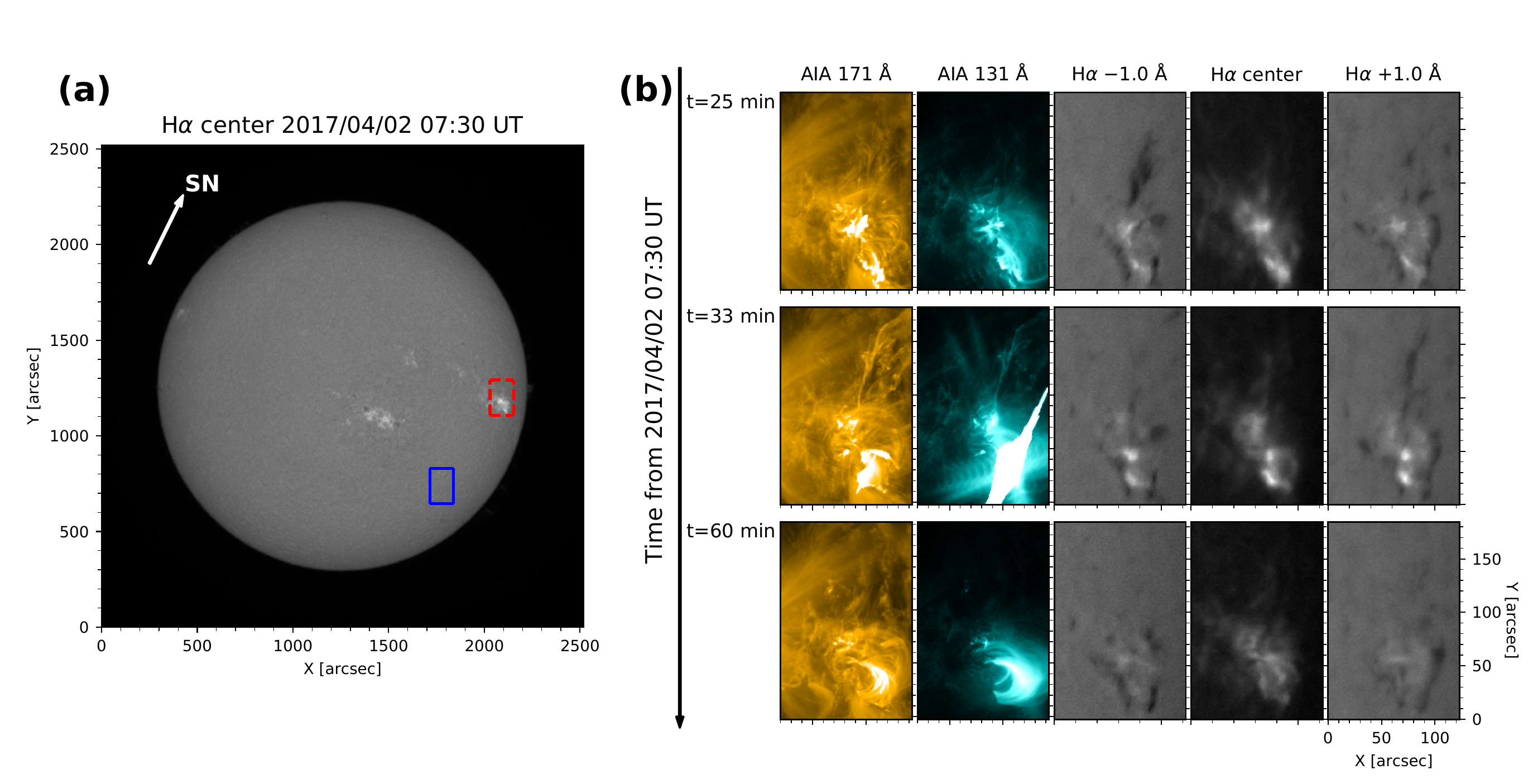}
\caption{Overview of event (2), which occurred on 2017 April 2. (a) A full-disk solar image at the H$\alpha$-center wavelength observed by SMART/SDDI. The red dashed region is a target region (TR), and the blue solid region is a quiet region (QR). Celestial north is up, and west is to the right. The direction of solar north is shown by the white arrow labeled as “SN”. (b) The time evolution of the M5.3 flare on 2017 April 2. The field of view in each panel corresponds to the TR in Figure \ref{IM_M53} (a). In each row,
images at the times t=25 min, t=33 min and t=60 min measured from 07:30 UT on 2017 April 2 are shown from top to bottom. In each column, the images AIA 171~{\AA}, AIA 131~{\AA}, H$\alpha-1.0$~{\AA}, H$\alpha$ center and H$\alpha+1.0$~{\AA} are shown from left to right. }
\label{IM_M53}
\end{figure}

Figure \ref{IM_M53} shows an overview of event (2). In this event, an M5.3 flare occurred on 2017 April 2 07:30 UT in NOAA 12644 near the solar limb [the red dashed region in Figure \ref{IM_M53} (a)]. This flare was associated with a filament eruption [Figure \ref{IM_M53} (b)]. After the \textit{GOES} peak time of this flare, a clear cusp-like structure was observed in the SDO/AIA image [Figure \ref{IM_M53} (b), t=60 min].

Figure \ref{fig:2017_0402_M5_3} shows the result of the Sun-as-a-star analysis of the H$\alpha$ spectra of this M5.3 flare. In the dynamic spectrum [Figure \ref{fig:2017_0402_M5_3} (a)], long-lasting emission appears near the H$\alpha$ line center. This emission extends up to $\pm2$ {\AA}.
Corresponding to the filament eruption on the disk, a blue-shifted absorption appears in the dynamic spectrum. However, even though the spatial scales of the eruptions are the same, $\sim100$ Mm for both event (1) and event (2), the strength of the absorption in event (2) is weaker than that in event (1). This would be the result of cancellation between the emission due to the flare and the absorption by the erupted plasma.
In this event, the filament erupted almost perpendicular to the line-of-sight (LOS) direction, with a very small LOS velocity. When the LOS velocity is very small, the filament appears near the H$\alpha$ line center, which is where the emission due to the flare appears, and absorption from the filament can cancel some of the emission from the flare.
Moreover, near the H$\alpha$ line center, where the background H$\alpha$ intensity is very small, the degree of absorption from the filament is also decreased dramatically compared to cases in which the filament appears in the far wing of the H$\alpha$ line.
Therefore, we also suggest that filament components are not dominant compared to the flare in this case, probably because the filament did not have a large contrast compared to the background due to its small LOS velocity (or possibly small density).
In addition, the $\Delta \mathrm{EW}_{\mathrm{H}\alpha\pm3.0\text{\AA}}$ does not show any absorption. The flaring amplitude is $\sim4\times10^{-4}$ {\AA}.
The time evolution of the $\Delta \mathrm{EW}_{\mathrm{H}\alpha\pm3.0\text{\AA}}$ is similar to the \textit{GOES} SXR light curve; for example, their time scales are comparable in each of the impulsive rising and gradual decay phases and their peaks are almost simultaneous [Figure \ref{fig:2017_0402_M5_3} (b)].
The similarity of the H$\alpha$ and the \textit{GOES} SXR light curves is also reported in the Sun-as-a-star analysis of a flare by \citet{NamekataETAL2022c}.

\subsection{Event (3), an M8.1 flare on 2017 September 8}
\begin{figure}
\centering
\includegraphics[width=14cm]
{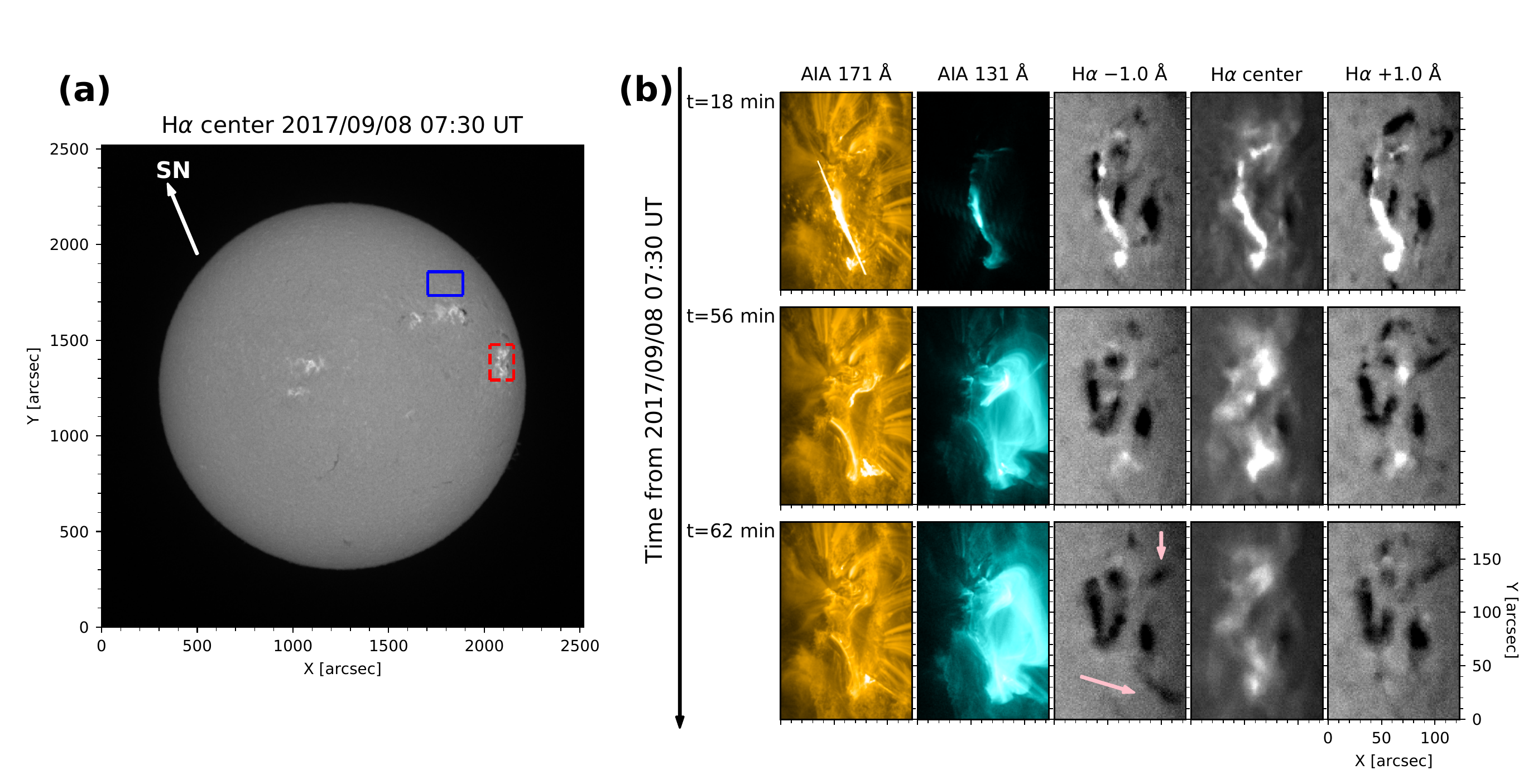}
\caption{Overview of event (3), which occurred on 2017 September 8. (a) A full-disk solar image at the H$\alpha$-center wavelength observed by SMART/SDDI. Celestial north is up, and west is to the right. The red dashed is a target region (TR), and the blue solid region is a quiet region (QR). The direction of solar north is shown by the white arrow labeled “SN”. (b) The time evolution of the M8.1 flare on 2017 September 8. The field of view in each panel corresponds to the TR in Figure \ref{IM_M81} (a). In each row, images at the times t=18 min, t=56 min and t=62 min measured from at 07:30 UT on 2017 September 8 are show from top to bottom. In each column, the images AIA 171~{\AA}, AIA 131~ {\AA}, H$\alpha-1.0$~{\AA}, H$\alpha$ center and H$\alpha+1.0$~{\AA} are shown from left to right. The pink arrows indicate plasma eruptions associated with the second brightening.  }
\label{IM_M81}
\end{figure}

Figure \ref{IM_M81} shows an overview of event (3). In this event, an M8.1 class flare occurred on 2017 September 8 07:30 UT in NOAA 12673 near the solar limb [the red dashed region in Figure \ref{IM_M81} (a)]. In the decay phase of the flare, a second brightening occurred. The first brightening was associated with a long ribbon-like structure and strong line broadening [Figure \ref{IM_M81} (b) t=18 min], while the second brightening was also associated with two small patches [Figure \ref{IM_M81} (b) t=56, 60 min]. The second brightening was followed by a filament eruption [Figure \ref{IM_M81} (b), t=60 min].

Figure \ref{fig:2017_0908} shows the result of the Sun-as-a-star analysis of the H$\alpha$ spectra of this M8.1 flare. In the dynamic spectrum [Figure \ref{fig:2017_0908} (a)], two-step emission appears near the H$\alpha$ line center. The first emission has strong line broadening, while the second emission exhibits weaker line broadening weaker than the first one. These emissions in the dynamic spectrum reflect features in the spatially resolved H$\alpha$ images [Figure\ref{IM_M81} (b)]. 
As shown in the spatially resolved H$\alpha$ image [Figure \ref{IM_M81} (b)] a plasma eruption was indeed occurring, but the dynamic spectrum does not show clear absorption.
This is probably because the filament erupted almost perpendicular to the LOS direction, as in event (2).
The $\Delta \mathrm{EW}_{\mathrm{H}\alpha\pm3.0\text{\AA}}$ does not show any absorption, on the other hand, it clearly shows two-step brightening (Figure \ref{fig:2017_0908} (b)). The flaring amplitudes are $\sim4\times10^{-4}$ {\AA} for the first brightening and $\sim2\times10^{-4}$ {\AA} for the second.

\subsection{Event (4), an M1.1 flare on 2021 April 19 - 20}
\begin{figure}
\centering
\includegraphics[width=14cm]
{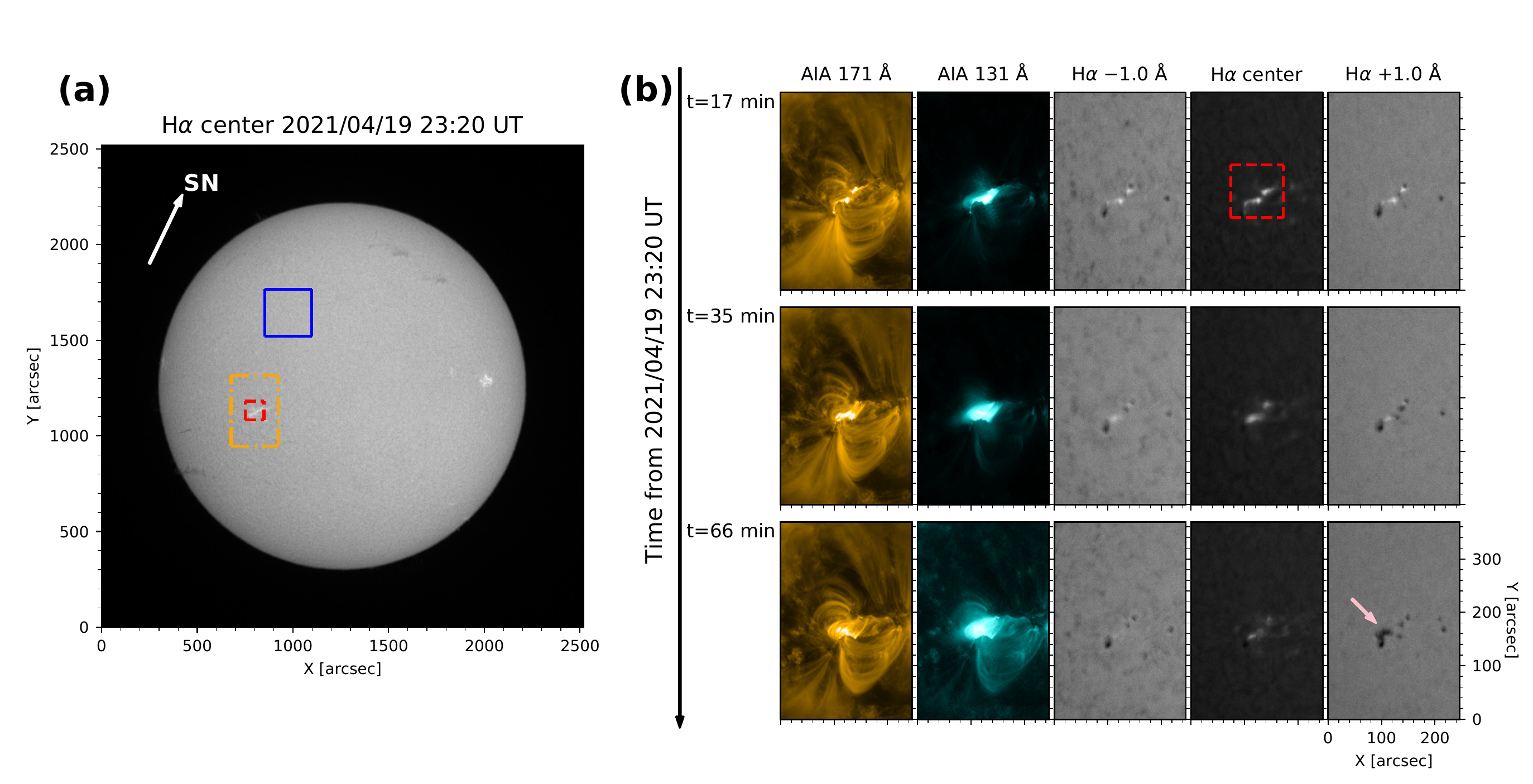}
\caption{Overview of event (4), which occurred on 2021 April 19-20. (a)A full disk solar image at the H$\alpha$-center wavelength observed by SMART/SDDI. Celestial north is up, and west is to the right. The orange dashdot region corresponds to the images in Figure \ref{IM_M11} (b), and the inner red dashed region is a target region (TR). The blue solid region is a quiet region (QR). The direction of solar north is shown by the white arrow labeled “SN”. (b) The time evolution of the M1.1-class flare on 2021 April 19. The field of view in each panel corresponds to the orange dashdot region in Figure \ref{IM_M11} (a). In each row, images at the times t=17 min, t=35 min and t=66 min measured from 23:30 UT on 2021 April 19 are shown from top to bottom. In each column, the images AIA 171~{\AA}, AIA 131~ {\AA}, H$\alpha-1.0$~{\AA}, H$\alpha$ center and H$\alpha+1.0$~{\AA} are shown from left to right. The red dashed region indicated in H$\alpha$ center image at t=17 min is a target region (TR). The pink arrow indicates plasma downflow during the decay phase of this M1.1 flare. }
\label{IM_M11}
\end{figure}

Figure \ref{IM_M11} shows an overview of event (4). In this event, an M1.1 class flare occurred on 2021 April 19 23:20 UT in NOAA 12816 near the disk center [the red dashed region in Figure \ref{IM_M11} (a),(b)]. 
Although this flare was not followed by any H$\alpha$ plasma eruption, plasma downflows were observed in H$\alpha$ red-wing images during the decay phase of this flare. 
Prior to these downflows, a hot loop was observed in the AIA 171 {\AA} and 131 {\AA} images at the same position [Figure \ref{IM_M11} (b)]. In a typical solar flare occurring on the solar disk, after the appearance of hot flare loops, cooled H$\alpha$ loops (“post-flare loops”) are observed as weak emissions at the H$\alpha$ line center. Moreover, cooled material moves downward along the post-flare loops (“coronal rain”), which is observed as absorption in the red wing of the H$\alpha$ line \citep[e.g.,][]{JingETAL2016}. 
Even in this event, the downflows in the flare-decay phase
of event (4) may be coronal rain from the post-flare loop. However, since we cannot resolve the H$\alpha$ loop that is expected to exist in the flare-decay phase, there is another interpretation of these downflows (e.g., partial drainage of ejected hot plasmas). 

Figure \ref{fig:2021_0420} shows the result of Sun-as-a-star analysis of the H$\alpha$ spectra of this M1.1 flare. In the dynamic spectrum [Figure \ref{fig:2021_0420} (a)], emission with a clear red asymmetry appears near the H$\alpha$ line center. When this emission has almost decayed, a red-shifted absorption appears. This absorption corresponds to the plasma downflows, which may be coronal rain from the post-flare loop occurring during the decay phase of the flare [Figure \ref{IM_M11} (b)]. The $\Delta \mathrm{EW}_{\mathrm{H}\alpha\pm3.0\text{\AA}}$ also exhibits the absorption after the peak of the flare [Figure \ref{fig:2021_0420} (b)]. The flaring amplitude is $\sim1.5\times10^{-4}$ {\AA}, and the amplitude of the absorption is $\sim0.5\times10^{-4}$ {\AA}.
\subsection{Event (5), a filament eruption on 2016 November 5}
\begin{figure}
\centering
\includegraphics[width=14cm]
{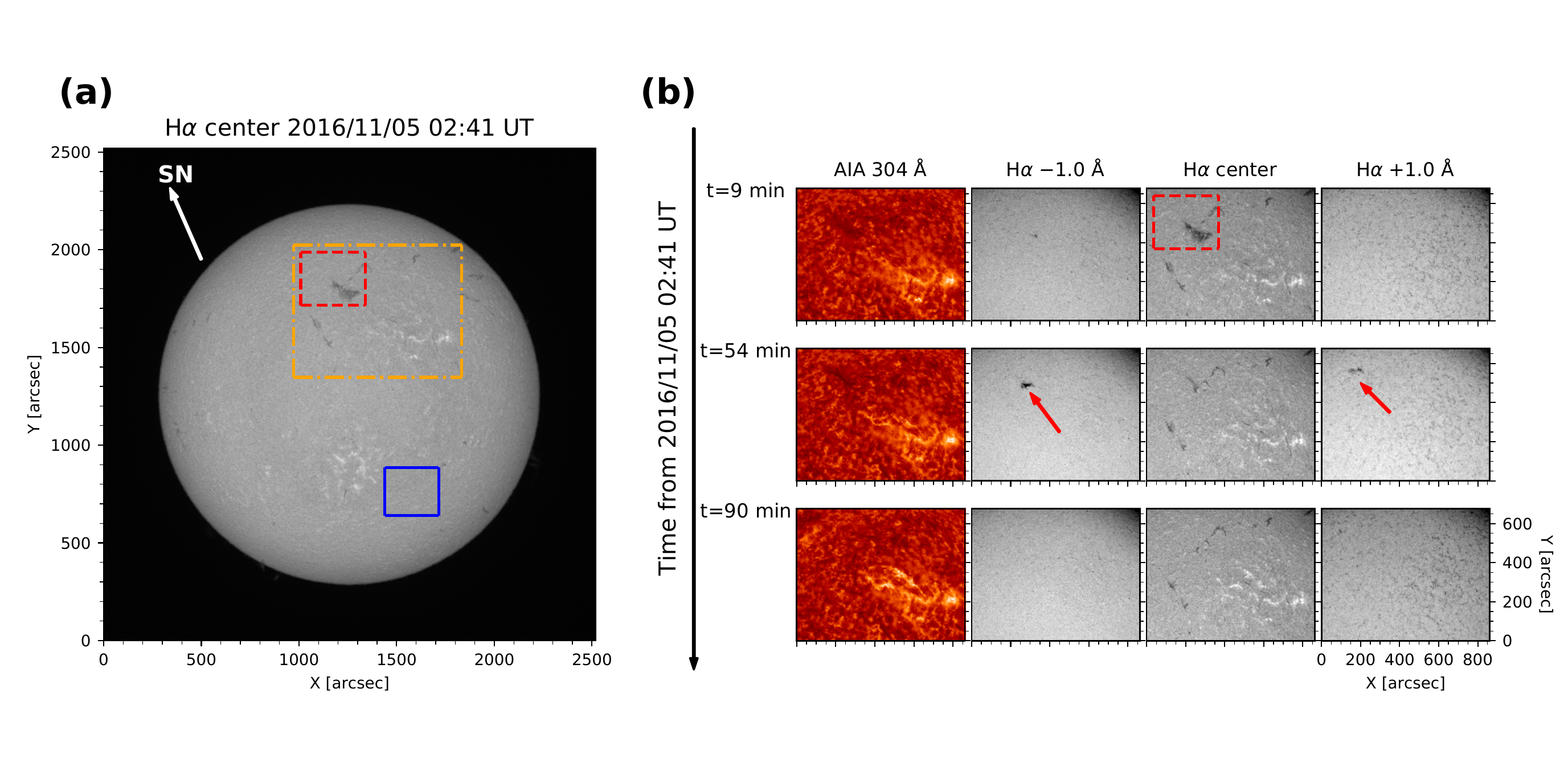}
\caption{Overview of event (5), which occurred on 2016 November 5. (a) A full-disk solar image at the H$\alpha$-center wavelength observed by SMART/SDDI. Celestial north is up, and west is to the right. The orange dashdot region corresponds to the images in Figure \ref{IM_FE1} (b), and the blue solid region is a quiet region (QR). The direction of solar north is shown by the white arrow labeled “SN”. (b) The time evolution of the filament eruption on 2016 November 5. The field of view in each panel corresponds to the orange dashdot region in Figure \ref{IM_FE1} (a). In each row,
images at the times t=9 min, t=54 min and t=90 min measured from 02:41 UT on 2016 November 5 are shown from top to bottom. In each column, the images AIA 304~{\AA}, H$\alpha-1.0$~ {\AA}, H$\alpha$ center and H$\alpha+1.0$~{\AA} are shown from left to right. The red dashed region indicated in the H$\alpha$ center image at t=9 min is a target region (TR). The red arrow in the H$\alpha-1.0$~{\AA} image at t=54 min indicates the upward motion of the filament eruption. The red arrow in the H$\alpha+1.0$~{\AA} image at t=54 min indicates the plasma downflow associated with the filament eruption.}
\label{IM_FE1}
\end{figure}

Figure \ref{IM_FE1} shows an overview of event (5). 
This event was analyzed from spatially-resolved viewpoints by \citet{SekiETAL2017}.
In this event, a filament eruption occurred on 2016 November 5 02:41 UT in the solar disk [the orange dashdot region in Figure \ref{IM_FE1} (a)]. The dark filament was observed in the H$\alpha$ image before the eruption. In connection with the filament eruption, a two-ribbon flare also occurred, but it was distant from the filament-eruption region. 
In order to model the phenomena that associated with only a filament eruption, we integrated H$\alpha$ spectra over the target region (TR) that included the filament eruption alone [the red dashed region in Figure \ref{IM_FE1} (a), (b)]. 

In events (5), (6), and (7), dark filaments were visible in the H$\alpha$ spectra before the eruptions. The contribution from the disappearance of such a dark filament to a Sun-as-a-star H$\alpha$ spectrum has not previously been investigated.
In event (6) and event (7), the filament and the two-ribbon flare were co-located. Thus it was difficult to separate the contributions from two-ribbon flares and from filament eruptions (the disappearance of the dark filament) to the formation of the Sun-as-a-star H$\alpha$ spectra in these two events. 
On the other hand, in event (5), the filament eruption occurred away from the two-ribbon flare (Figure \ref{IM_FE1}).
By using this spatial separation between the filament and the flare in event (5), we have been able to perform the Sun-as-a-star analysis for the filament eruption alone in event (5) and to investigate its contribution to the formation of the Sun-as-a-star H$\alpha$ spectra.

Figure \ref{fig:2016_1105} shows the result of the Sun-as-a-star analysis of H$\alpha$ spectra of this filament eruption. In the dynamic spectrum
[Figure \ref{fig:2016_1105} (a)], the emission with shifted absorptions appears near the H$\alpha$ line center. 
Since the H$\alpha$ spectra were integrated over the target region (TR) that included the filament eruption alone, the emission seen in the H$\alpha$ line center is not due to the distant flare ribbon but instead is produced because the dark filament, which absorbed the background radiation at H$\alpha$ line center  before the eruption, had disappeared from the line center. The blue- and red-shifted absorptions correspond, respectively, to upward motions of the plasma and to partial drainage associated with the filament eruption [Figure \ref{IM_FE1} (b)]. The $\Delta \mathrm{EW}_{\mathrm{H}\alpha\pm3.0\text{\AA}}$. also shows absorption and emission [Figure \ref{fig:2016_1105} (b)], the amplitudes of which are both $\sim1\times10^{-4}$ {\AA}.
The quantity $\Delta \mathrm{EW}_{\mathrm{H}\alpha\pm0.5\text{\AA}}$ is also calculated only over the range that includes the emission. The emission in $\Delta \mathrm{EW}_{\mathrm{H}\alpha\pm0.5\text{\AA}}$ appears at the almost same time as the absorption in $\Delta \mathrm{EW}_{\mathrm{H}\alpha\pm3.0\text{\AA}}$.
This simultaneity is consistent with the depiction that the emission comes from the disappearance of the dark filament.

How strong is the emission due to the disappearance of the dark filament compared to that from the two-ribbon flare? To address this question, we also calculated the differenced equivalent width for the two-ribbon flare alone. As shown in Figure \ref{fig:2016_1105_FE_FL}, brightening from the disappearance of the dark filament is comparable to that from the two-ribbon flare. This result means that the contribution to the formation of a Sun-as-a-star H$\alpha$ spectrum from apparent emission by filament disappearance cannot be negligible compared to that from emission by a two-ribbon flare.

\subsection{Event (6), a filament eruption and a two-ribbon flare on 2017 February 19}
\begin{figure}
\centering
\includegraphics[width=14cm]
{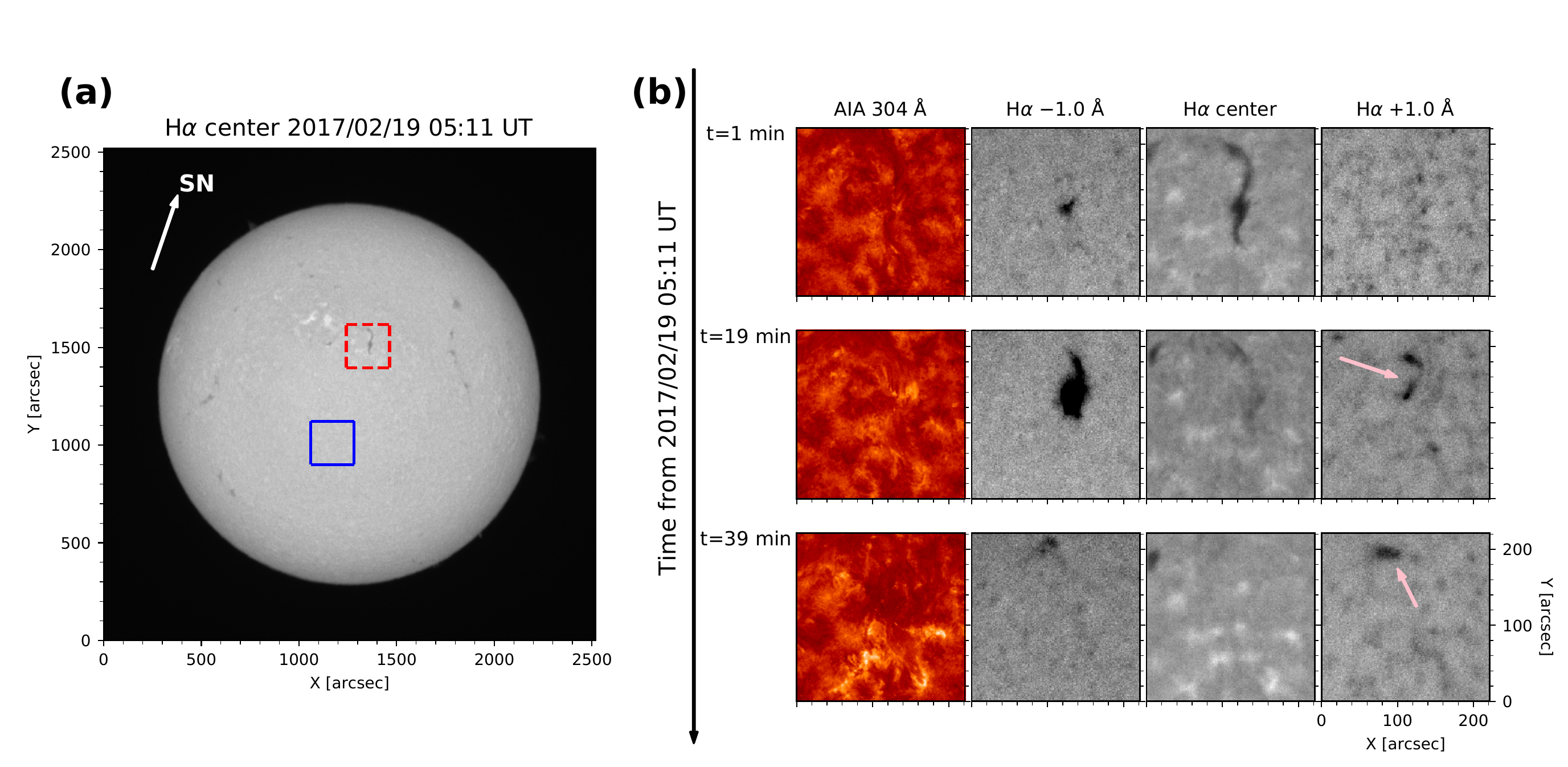}
\caption{Overview of event (6), which occurred on 2017 February 19. (a) A full-disk solar image at the H$\alpha$-center wavelength observed by SMART/SDDI. Celestial north is up, and west is to the right. The red dashed region is a target region (TR), and the blue solid region is a quiet region (QR). The direction of solar north is shown by the white arrow labeled “SN”. (b) The time evolution of the filament eruption on 2017 February 19. The field of view in each panel corresponds to the TR in Figure \ref{IM_FE2} (a). In each row,
images at the times t=1 min, t=19 min and t=39 min measured from 05:11 UT on 2017 February 19 are shown from top to bottom. In each column, the images AIA 304~{\AA}, H$\alpha-1.0$~ {\AA}, H$\alpha$ center and H$\alpha+1.0$~{\AA} are shown from left to right. The pink arrows indicate plasma downflows associated with the filament eruption.}
\label{IM_FE2}
\end{figure}

Figure \ref{IM_FE2} shows an overview of event (6). In this event, a filament eruption occurred on 2017 February 19 05:11 UT co-located with a two-ribbon flare [the red dashed region in Figure \ref{IM_FE2} (a)].

Figure \ref{fig:2017_0219} shows the result of the Sun-as-a-star analysis of the H$\alpha$ spectra of this filament eruption and two-ribbon flare. In the dynamic spectrum [Figure \ref{fig:2017_0219} (a)], emission with shifted absorptions appears near the H$\alpha$ line center. In the same way as event (5), the blue- and red-shifted absorptions correspond, respectively, to the upward motions of the plasma and to partial drainage associated with the filament eruption [Figure \ref{IM_FE2} (b)]. 
Before the fast blue shifted absorption, a slow blue-shifted absorption appears in the dynamic spectrum. This slow absorption probably comes from the slow upward motions of the filaments before the eruption that was reported in many previous studies from spatially-resolved viewpoints \citep{SterlingMoore2004, IsobeTripathi2006, SekiETAL2017, SekiETAL2019a} .
We also note that the fast blue-shifted absorption is divided into two stripes. The first stripe ends at about $-4$ {\AA} ($\sim-200$ km s$^{-1}$), while the second stripe is decelerated and connected to the red shifted absorption. The first stripe comes from erupted blobs that were accelerated to about $-200$ km s$^{-1}$ and finally faded away in the H$\alpha$ images. The second stripe comes from erupted blobs that eventually fell back into the solar surface. 

In this event, we include the two-ribbon flare in the TR. The AIA 304 {\AA} image shows a two-ribbon flare relatively clearly. Its counterparts are also observed in the H$\alpha$ line-center image [Figure \ref{IM_FE2} (b)]. 
Therefore, the emission near the H$\alpha$ line center is a superposition of contributions from the two-ribbon flare and from the disappearance of the dark filament.
The quantity $\Delta \mathrm{EW}_{\mathrm{H}\alpha\pm4.0\text{\AA}}$ also shows the absorption and emission [Figure \ref{fig:2017_0219} (b)], the amplitudes of which are $\sim3\times10^{-4}$ {\AA} and $\sim1\times10^{-4}$ {\AA}, respectively. The quantity $\Delta \mathrm{EW}_{\mathrm{H}\alpha\pm0.5\text{\AA}}$ is also calculated only over the range that includes the emission. The absorption in $\Delta \mathrm{EW}_{\mathrm{H}\alpha\pm4.0\text{\AA}}$ appears faster than the emission in $\Delta \mathrm{EW}_{\mathrm{H}\alpha\pm0.5\text{\AA}}$. 
The faster appearance of this absorption may be the result of the slow upward motions of the filament before the fast eruption. 
\subsection{Event (7), a filament eruption and a two-ribbon flare on 2017 April 23}
\begin{figure}
\centering
\includegraphics[width=14cm]
{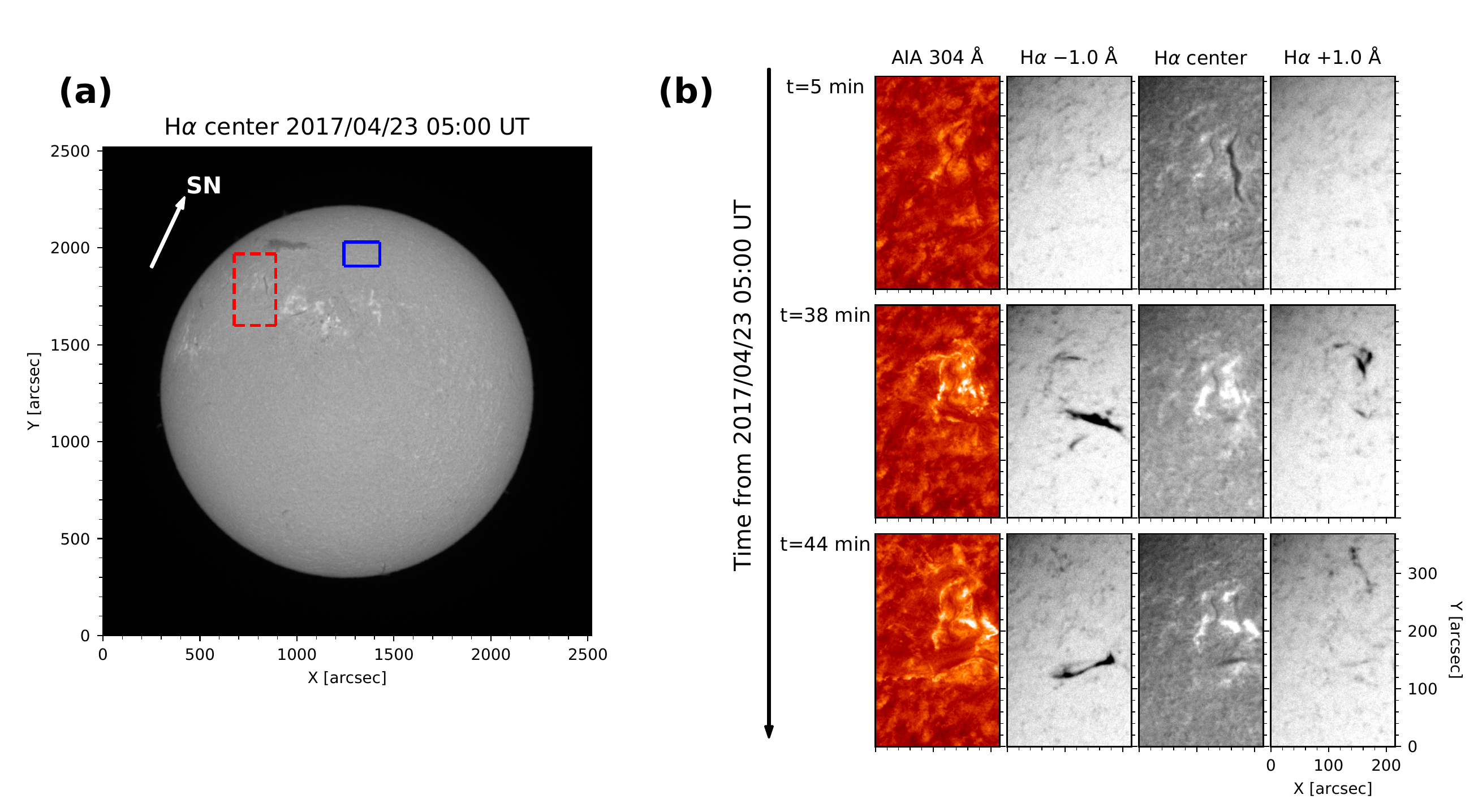}
\caption{Overview of event (7), which occurred on 2017 April 23. (a) A full-disk solar image at the H$\alpha$-center wavelength observed by SMART/SDDI. Celestial north is up, and west is to the right. The red dashed region is a target region (TR), and the blue solid region is a quiet region (QR). The direction of solar north is shown by the white arrow labeled “SN”. (b) The time evolution of the filament eruption on 2017 April 23. The field of view in each panel corresponds to the TR in Figure \ref{IM_FE3} (a). In each row,
images at the times t=5 min, t=38 min and t=44 min measured from 05:00 UT on 2017 April 23 are shown from top to bottom. In each column, the images AIA 304~{\AA}, H$\alpha-1.0$~ {\AA}, H$\alpha$ center and H$\alpha+1.0$~{\AA} are shown from left to right. }
\label{IM_FE3}
\end{figure}

Figure \ref{IM_FE3} shows an overview of event (7). In this event, a filament eruption occurred on 2017 April 23 05:00 UT co-located with a two-ribbon flare [the red dashed region in Figure \ref{IM_FE3} (a)].

Figure \ref{fig:2017_0423} shows the result of the Sun-as-a-star analysis of the H$\alpha$ spectra of this filament eruption. In the dynamic spectrum [Figure \ref{fig:2017_0423} (a)], the emission with shifted absorptions appears near the H$\alpha$ line center. In the same way as in event (5) and event (6), the blue- and red-shifted absorptions correspond, respectively, to the upward motions of the plasma and partial drainage associated with the filament eruption [Figure \ref{IM_FE3} (b)]. In this event, we include the two-ribbon flare in the TR, as for event (6). The AIA 304 {\AA} image shows the two-ribbon flare relatively clearly, and its counterpart also appears clearly in the H$\alpha$ line-center image [Figure \ref{IM_FE3} (b)]. 
The emission near the H$\alpha$ line center is therefore a superposition of contributions from the two-ribbon flare and from the disappearance of the dark filament, as in event (6).
The quantity $\Delta \mathrm{EW}_{\mathrm{H}\alpha\pm3.0\text{\AA}}$ also exhibits the absorption and emission [Figure \ref{fig:2017_0423} (b)], the amplitudes of which are $\sim3\times10^{-4}$ {\AA} and $\sim1\times10^{-4}$ {\AA}, respectively. 
The quantity $\Delta \mathrm{EW}_{\mathrm{H}\alpha\pm0.5\text{\AA}}$ is also calculated only over the range that includes the emission. 
The time evolution of the $\Delta \mathrm{EW}_{\mathrm{H}\alpha\pm0.5\text{\AA}}$ is similar to the \textit{GOES} SXR light curve. This similarity is consistent with the clear two-ribbon flare in the H$\alpha$ line. 

\subsection{Event (8), a prominence eruption on 2017 June 19 }
\begin{figure}
\centering
\includegraphics[width=14cm]
{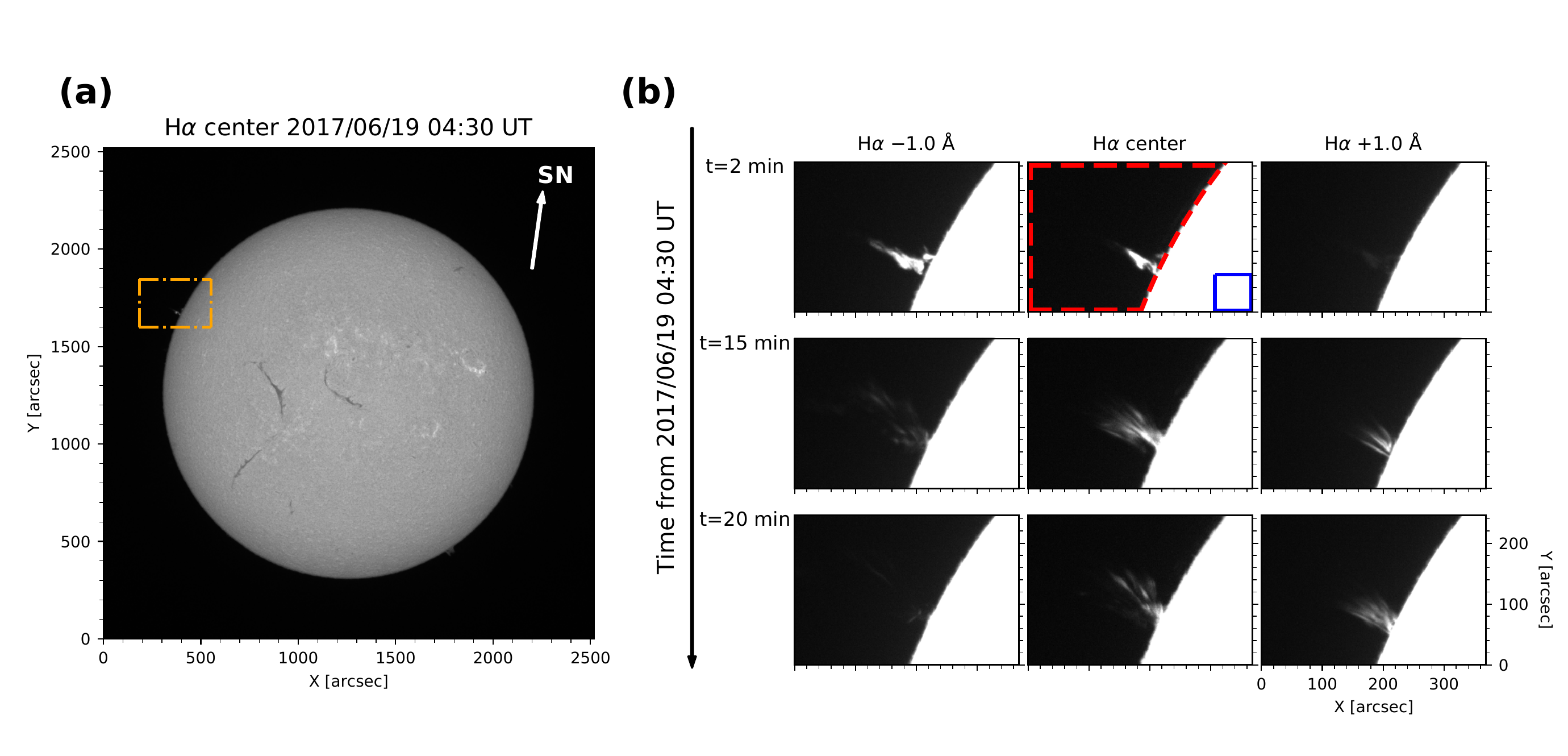}
\caption{Overview of event (8), which occurred on 2017 June 19. (a)A full disk solar image at the H$\alpha$-center wavelength observed by SMART/SDDI. Celestial north is up, and west is to the right. The orange dashdot region corresponds to the images in Figure \ref{IM_PE1} (b). The direction of solar north is shown by the white arrow labeled “SN”. (b) The time evolution of the filament eruption on 2017 June 19. The field of view in each panel corresponds to the orange dashdot region in Figure \ref{IM_PE1} (a). In each row,
images at the times t=2 min, t=15 min and t=20 min measured from 04:30 UT 2017 June 19 are show from top to bottom. In each column, the images H$\alpha-1.0$~{\AA}, H$\alpha$ center and H$\alpha+1.0$~{\AA} are shown from left to right. The red dashed region in the H$\alpha$ center image at t=2 min is a target region (TR), and the blue solid region is a quiet region (QR). }
\label{IM_PE1}
\end{figure}

Figure \ref{IM_PE1} shows an overview of event (8). In this event, a prominence eruption to the outside of the solar limb occurred on 2017 June 19 04:30 UT [the orange dashdot region in Figure \ref{IM_PE1} (a)].
We integrated H$\alpha$ spectra over the region outside of the solar disk [the red dashed region in Figure \ref{IM_PE1} (b)]. The H$\alpha$ flare was not seen from Earth, so the footpoints of this event must be on the backside of the solar limb.

Figure \ref{fig:2017_0619} shows the result of the Sun-as-a-star analysis of the H$\alpha$ spectra of this prominence eruption. 
Note that there are no pre-event data before 2017 June 19 04:30 UT because calibration frames were being taken. We have therefore used post-event data (after this event) in place of the pre-event data in analyzing this event.
In the dynamic spectrum [Figure \ref{fig:2017_0619} (a)], strongly shifted emission appears, which transforms from a blue shift to a red shift. In this event, the prominence erupted beyond the solar limb, with the LOS velocity approaching Earth. Part of the erupted plasma subsequently fell into the solar surface, with the LOS velocity moving away from Earth [Figure \ref{IM_PE1} (b)]. The shifted emission corresponds to this series of plasma motions outside the solar limb.  The quantity $\Delta \mathrm{EW}_{\mathrm{H}\alpha\pm3.0\text{\AA}}$ also shows the emission [Figure \ref{fig:2017_0619} (b)], with an amplitude $\sim3\times10^{-4}$ {\AA}.

\subsection{Event (9), a prominence eruption on 2021 May 5}
\begin{figure}
\centering
\includegraphics[width=14cm]
{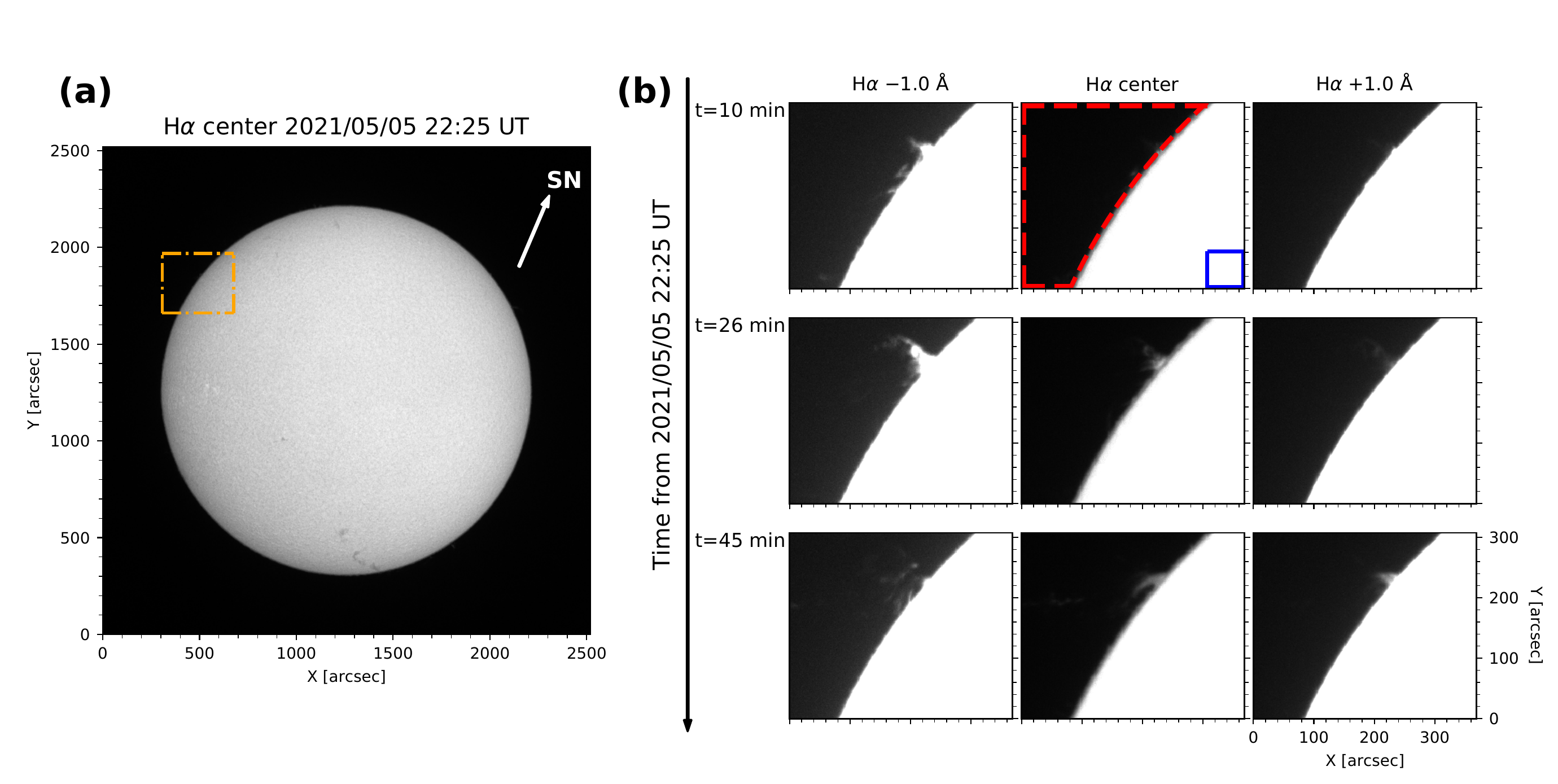}
\caption{Overview of event (9), which occurred on 2021 May 5. (a) A-full disk solar image at the H$\alpha$-center wavelength observed by SMART/SDDI. Celestial north is up, and west is to the right. The orange dashdot region corresponds to the images in Figure \ref{IM_PE2} (b). The direction of solar north is shown by the white arrow labeled “SN”. (b) The time evolution of the filament eruption on 2021 May 5. The field of view in each panel corresponds to the orange dashdot region in Figure \ref{IM_PE2} (a). In each row,
images at the times t=10 min, t=26 min and t=45 min measured from 22:25 UT on 2021 May 5 are shown from top to bottom. In each column, the images of H$\alpha-1.0$~{\AA}, H$\alpha$ center and H$\alpha+1.0$~{\AA} are shown from left to right. The red dashed region in the H$\alpha$ center image at t=10 min is a target region (TR), and the blue solid region is a quiet region (QR). }
\label{IM_PE2}
\end{figure}

Figure \ref{IM_PE2} shows an overview of event (9). In this event, a prominence eruption to the outside of the solar limb occurred on 2021 May 5 22:25 UT [the orange dashdot region in Figure \ref{IM_PE2} (a)].
We integrated H$\alpha$ spectra over the region outside of the solar disk [the red dashed region in Figure \ref{IM_PE2} (b)].
The H$\alpha$ flare ribbon is not seen from Earth, so the footpoints of this event must be on the backside of the solar limb, as for event (8).

Figure \ref{fig:2021_0505} shows the result of the Sun-as-a-star analysis of the H$\alpha$ spectra of this prominence eruption. 
In the dynamic spectrum [Figure \ref{fig:2021_0505} (a)], a blue-shifted emission first appears. When this emission approaches the H$\alpha$ line center, a second emission component appears, drifting from the blue wing to the red wing. In this event, the two successive prominence eruptions occurred, and both erupted with their LOS velocities approaching Earth. 
After the second eruption, a part of the prominence fell back into the solar surface, with the LOS velocity moving away from Earth. The successive emissions in the dynamic spectrum come from this series of plasma motions outside the solar limb. Corresponding to these successive emissions, the quantity $\Delta \mathrm{EW}_{\mathrm{H}\alpha\pm9.0\text{\AA}}$ for event (9) shows a two-step brightening [Figure \ref{fig:2021_0505} (b)], with emission amplitudes $\sim1\times10^{-4}$ {\AA} for the first one and $\sim2\times10^{-4}$ {\AA} for the second one.

\section{Summary and Implications to stellar observations
 \label{s:SD}}
\begin{deluxetable*}{ccccccc}
\label{table2}
\tablenum{2}
\tablecaption{Summary of the results of Sun-as-a-star analysis}
\tablewidth{0pt}
\tablehead{
\colhead{Event} &
\colhead{Event features} & \colhead{Features in Sun-as-a-star spectra} & 
\colhead{$\pm\Delta\lambda$ [{\AA}]}&
\colhead{Em. [{\AA}]}& 
\colhead{Abs. [{\AA}]}
}
\decimals
\startdata
(1) & flare + surge & 
Em., Shifted abs..&
$\pm3.0$
& $3\times10^{-4}$ & $2\times10^{-4}$\\
(2) & flare + eruption &
Em., Blue shifted abs. (weak)&
$\pm3.0$
& $4\times10^{-4}$ &  none\\
(3) & flare (two-step) + eruption &
Em. (two step) &
$\pm3.0$
& $4\times10^{-4}$, $2\times10^{-4}$& none\\
(4) & flare + coronal rain ?&
Em., Red shifted abs. &
$\pm3.0$
& $1.5\times10^{-4}$ & $0.5\times10^{-4}$\\
(5) & filament eruption &
Em., Shifted abs. &
$\pm3.0$
 & $1\times10^{-4}$ & $1\times10^{-4}$\\
(6) & two-ribbon flare + filament eruption &
 Em., Shifted abs. &
$\pm4.0$
 & $1\times10^{-4}$ 
 & $3\times10^{-4}$\\
(7) & two-ribbon flare + filament eruption &
 Em., Shifted abs.&
$\pm3.0$
 & $1\times10^{-4}$ 
 & $3\times10^{-4}$\\
(8) & prominence eruption &
Shifted em.&
$\pm3.0$
& $3\times10^{-4}$
& none\\
(9) & prominence eruption (two-step)&
Shifted em. (two step) &
$\pm9.0$
& $2\times10^{-4}$, $1\times10^{-4}$
& none\\
\enddata
\tablecomments{In this Table, “Em.” and “Abs.” represent emission and absorption, respectively.
In the columns labeled Em. [{\AA}] and Abs. [{\AA}], the approximate amplitudes of emissions and absorptions in differenced equivalent widths are shown, respectively. In the column labeled $\pm\Delta\lambda$ [{\AA}], we have summarized the integral ranges used for calculating the differenced equivalent widths in the columns of Em. [{\AA}] and Abs. [{\AA}].}
\end{deluxetable*}
We have performed Sun-as-a-star analyses of the H$\alpha$ spectra of the various active events on the Sun; specifically, flares ($\geq$ C8.0), filament eruptions with large spatial scales and eruptions of off limb prominences
. Our aim is to increase the number of samples beyond those provided by \citet{NamekataETAL2022a,NamekataETAL2022c}---who reported Sun-as-a-star analyses of a solar flare, a filament eruption, and a surge---and to clarify the correspondence between various active events and features seen in Sun-as-a-star H$\alpha$ spectra for future applications to stellar observations. We have summarized our results in Table \ref{table2}. All nine events analyzed in this study show emission relative to the pre-event level. The amplitudes of their differenced equivalent widths are all on the order of $10^{-4}$ {\AA}, regardless of the cause of the emission. On the other hand, the dynamic spectra show different features, depending on the cause of the emission: 
(i) Flares show emission at the H$\alpha$ line center with red asymmetry and line broadening as reported in a previous study.
(ii) Filament eruptions with and without flares show emission near the H$\alpha$ line center, together with blue-/red-shifted absorptions.
(iii) Eruptions of off limb prominences show blue-/red-shifted emissions. 
If spectral features similar to those of (i) $-$ (iii) are obtained in stellar observations, we will thus be able to identify the phenomena occurring
on those stars. We summarize below the details of the features in the Sun-as-a-star H$\alpha$ spectra.

\begin{figure}
\centering
\includegraphics[width=8cm]
{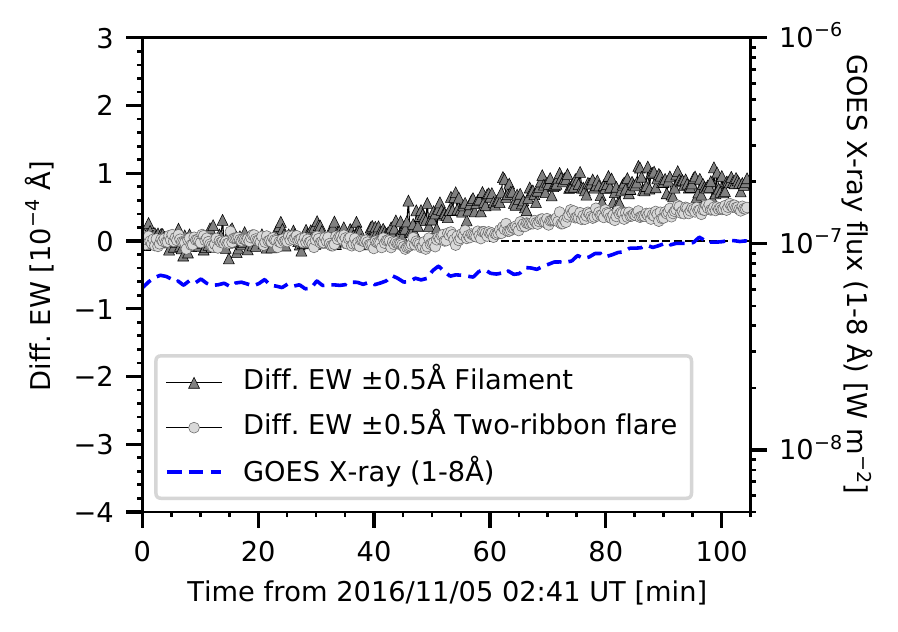}
\caption{The contributions to the differenced equivalent widths from the filament eruption and the two-ribbon flare. 
The differenced equivalent widths of H$\alpha\pm0.5$~{\AA} for the filament eruption are plotted as gray triangles and for the two-ribbon flare as lightgray circles. The \textit{GOES} X-ray (1$-$8{\AA}) flux is also plotted as a blue dashed line.
\label{fig:2016_1105_FE_FL}}


\end{figure}

\begin{itemize}
\item[(i)] Flares ($\geq$ C8.0) show emission near the H$\alpha$ line center, with red asymmetry and/or line broadening in their Sun-as-a-star H$\alpha$ spectra. \citet{NamekataETAL2022c} also reported that the Sun-as-a-star H$\alpha$ spectrum of an M4.2 solar flare near the disk center showed red asymmetry and line broadening, and they thus suggested that the nature of the chromospheric condensation can be seen even in the Sun-as-a-star spectra. We note that, although event (2) and event (3) are especially energetic events in our study, red asymmetry is not very clear in them. These two events occurred near the solar limbs, and the lack of red asymmetry can be explained as a projection effect, as reported in previous studies \citep[e.g.,][]{SvestkaETAL1962}.
The red asymmetry associated with flares can be interpreted as chromospheric condensation moving downward \citep[e.g.,][]{IchimotoKurokawa1984, AsaiETAL2012}. Therefore, assuming that material moves along a direction radial from the Sun, flares occurring near the solar limb should show a weak red asymmetry. In event (4), a red-shifted absorption appears in the Sun-as-a-star H$\alpha$ spectra when the emission due to the M1.1 flare almost decayed. This red-shifted absorption during the flare decay phase is the candidate of coronal rains from post-flare loops.

\item[(ii)] Filament eruptions with sufficient LOS velocity result in both emission at the H$\alpha$ line center and shifted absorptions in the Sun-as-a-star H$\alpha$ spectra.
Emissions from flares in Sun-as-a-star H$\alpha$ spectra have been reported previously by \citet{NamekataETAL2022a,NamekataETAL2022c}, but we found that additional apparent emission at the H$\alpha$ line center due to disappearance of a dark filament.
As shown in Figure \ref{fig:2016_1105_FE_FL}, such apparent emission can be comparable to emission due to a two-ribbon flare and it can contribute to the H$\alpha$ spectral-line formation.
Before fast eruptions, slow blue-shifted absorptions appear in the Sun-as-a-star H$\alpha$ spectra [clearly seen in event (6), Figure \ref{fig:2017_0219} (a)]. These absorptions probably come from slow motions of the filaments before the eruptions, as reported in many previous studies with spatially resolved images \citep{SterlingMoore2004,IsobeTripathi2006,SekiETAL2017,SekiETAL2019a}.
In addition to the blue shifted absorption due to upward motions of the filaments, red-shifted absorptions appear together with the emissions in the dynamic spectra. These red-shifted absorptions come from partial drainage of the erupting filaments.

\item[(iii)] Prominence eruptions that erupted outside the solar limb result in shifted emission in the Sun-as-a-star H$\alpha$ spectra. It is well known from spatially-resolved viewpoints, that when a prominence erupts beyond the solar limb, emission occurs in the H$\alpha$ spectra because of the absence of background light \citep{Parenti2014}. The prominences of event (8) and event (9) had velocities along the LOS direction, and the shifted emissions therefore appears in the Sun-as-a-star H$\alpha$ spectra.
We demonstrate for the first time how the H$\alpha$ spectra of prominence eruption can be seen in the Sun-as-a-star view. 
\end{itemize}

\begin{figure}
\centering
\includegraphics[width=8cm]
{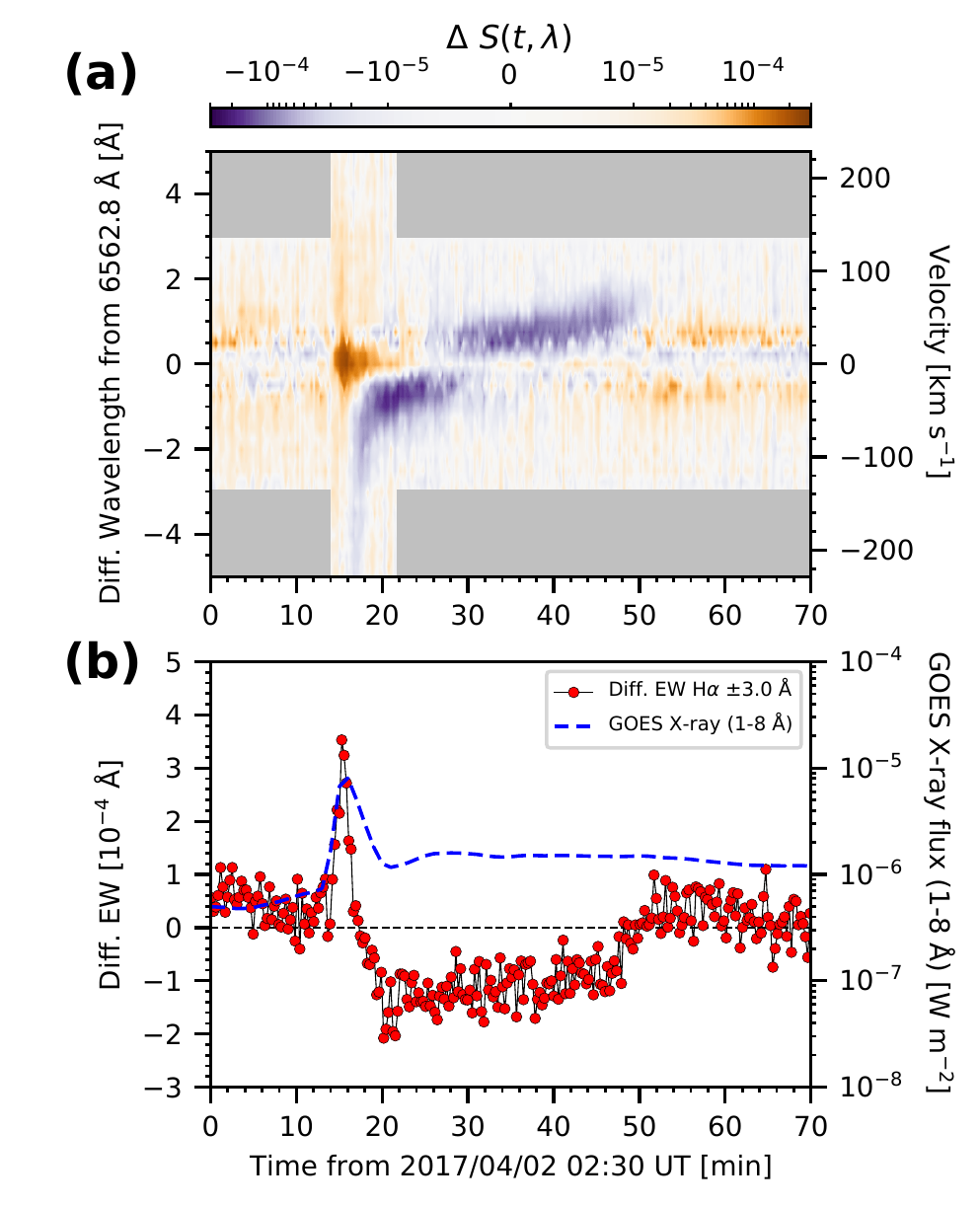}
\caption{(a) The time series of the Sun-as-a-star H$\alpha$ spectra (dynamic spectrum) and (b) the light curves of event (1), the C8.0 flare on 2017 April 2. In panel (a), the left vertical axis is the difference in wavelength from the H$\alpha$ line center in units of {\AA}, and the right vertical axis is the corresponding Doppler velocity in units of km s$^{-1}$. The horizontal axis is the time measured from 02:30 UT on 2017 April 2 in units of minutes. The color map shows $\Delta S(t,\lambda)$, where $\Delta S(t,\lambda)$ represents the ratio of the changes in the pre-event-subtracted spatially integrated H$\alpha$ spectra to the full-disk continuum (see Eq. \ref{dS}). 
The orange and purple regions correspond to emission and absorption, respectively.
In the gray regions, there is no available data.  
Panel (b) shows the differenced equivalent widths of H$\alpha\pm3.0$~{\AA} as red circles and the GOES soft X-ray (1$-$8~{\AA}) flux plotted a blue dashed line. The horizontal axis is the same as that in panel (a). The horizontal dashed line represents the zero level of differenced equivalent width.
\label{fig:2017_0402_C}}
\end{figure}

\begin{figure}
\centering
\includegraphics[width=8cm]
{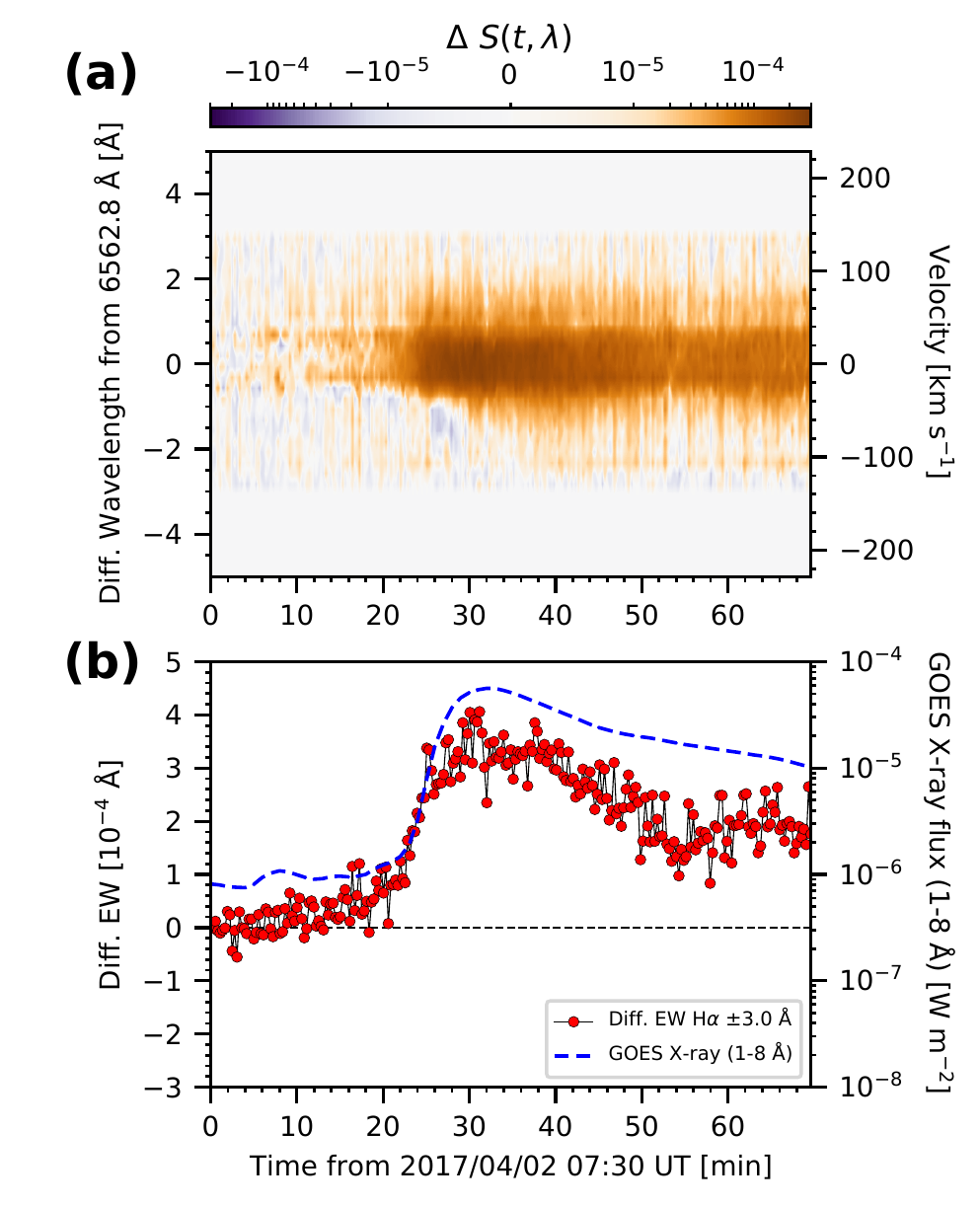}
\caption{The same as Figure \ref{fig:2017_0402_C}, but for event (2), the M5.3 flare on 2017 April 2. We note that only H$\alpha\pm3.0$~{\AA} data were available for performing the Sun-as-a-star analysis in this event.
\label{fig:2017_0402_M5_3}}
\end{figure}

\begin{figure}
\centering
\includegraphics[width=8cm]
{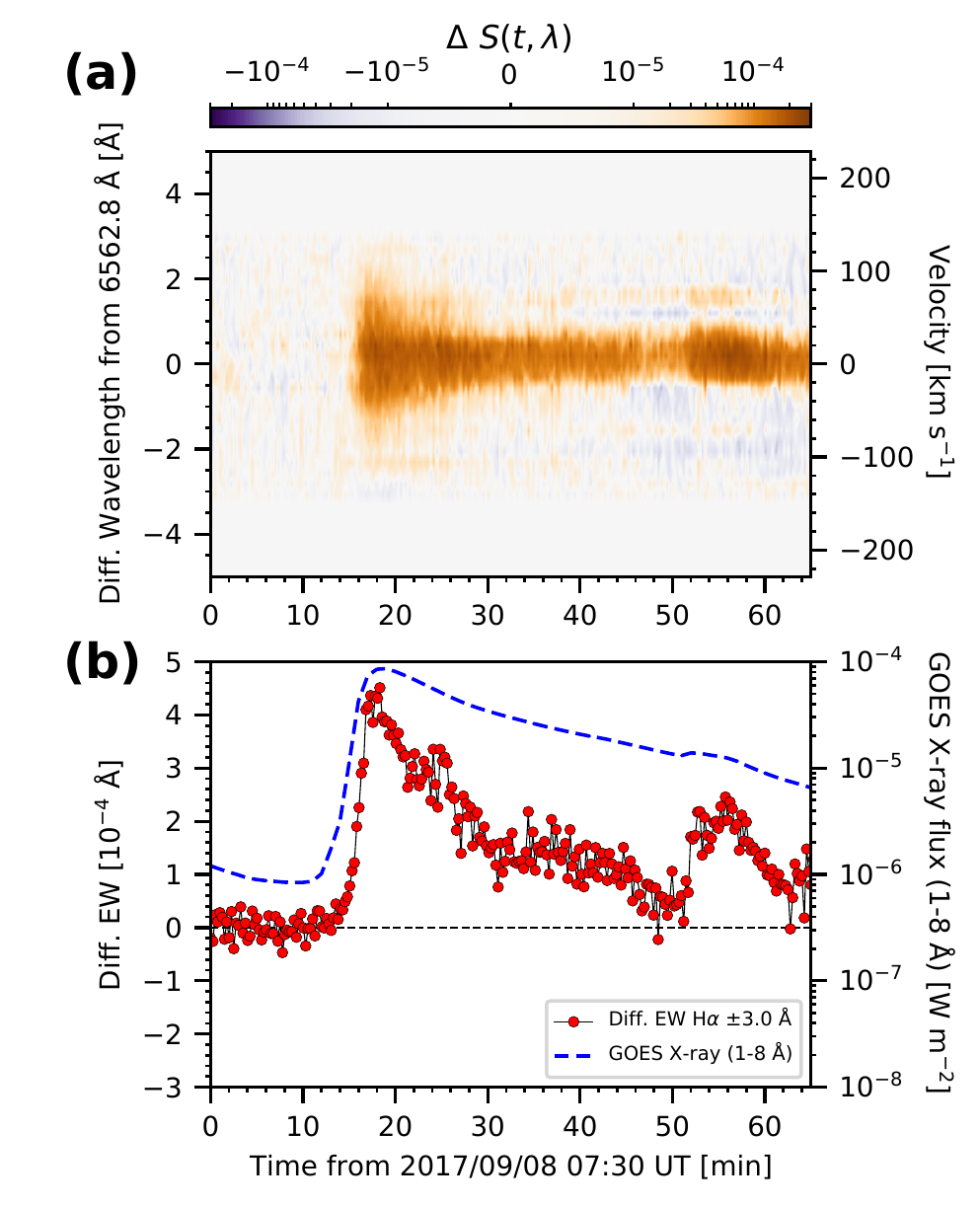}
\caption{The same as Figure \ref{fig:2017_0402_C}, but for event (3), the M8.1 flare on 2017 September 8. We note that only H$\alpha\pm3.0$~{\AA} data were available for performing the Sun-as-a-star analysis in this event.
\label{fig:2017_0908}}
\end{figure}

\begin{figure}
\centering
\includegraphics[width=8cm]
{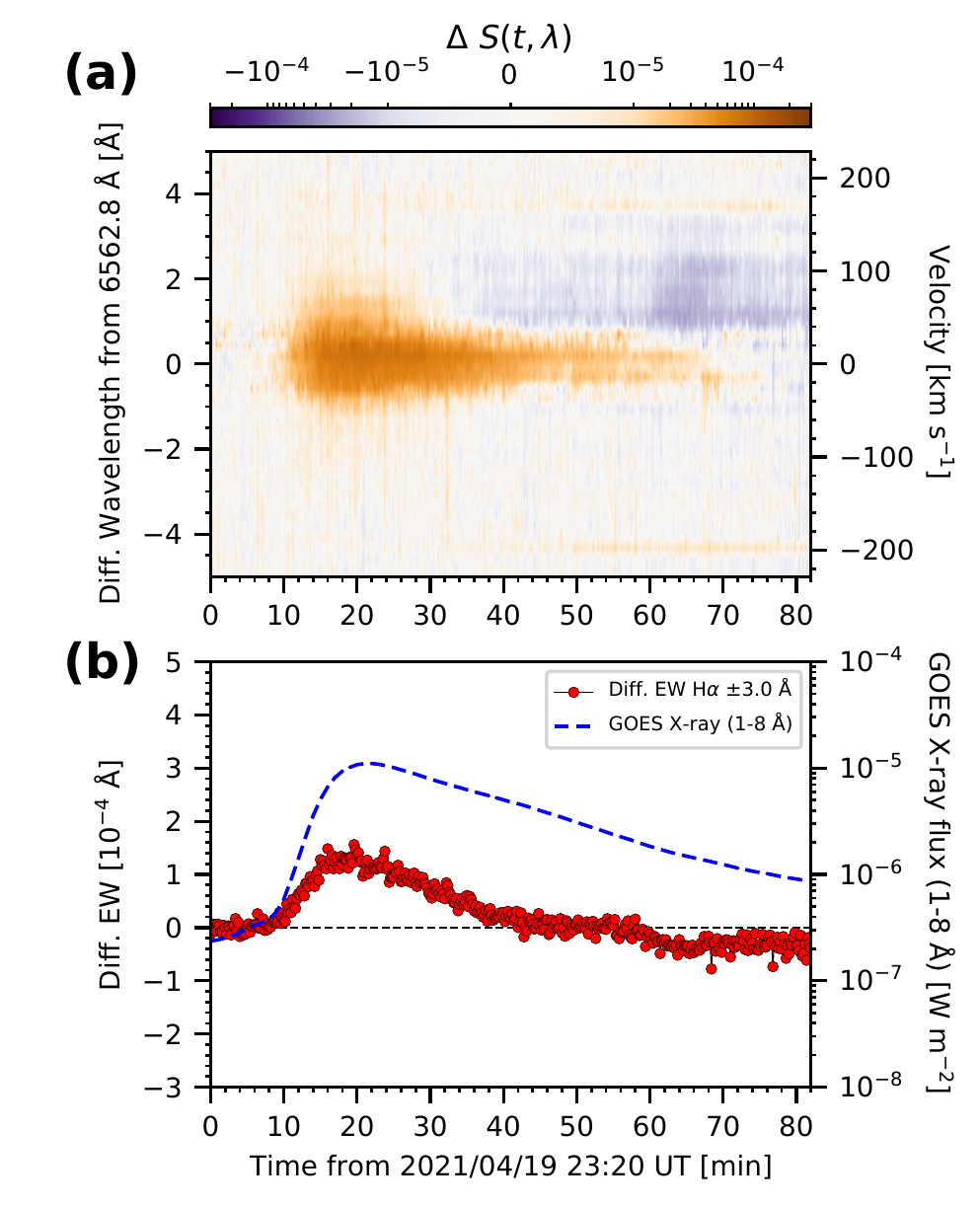}
\caption{The same as Figure \ref{fig:2017_0402_C}, but for event (4), the M1.1 flare on 2021 April 19 to 20. 
\label{fig:2021_0420}}


\end{figure}

\begin{figure}
\centering
\includegraphics[width=8cm]
{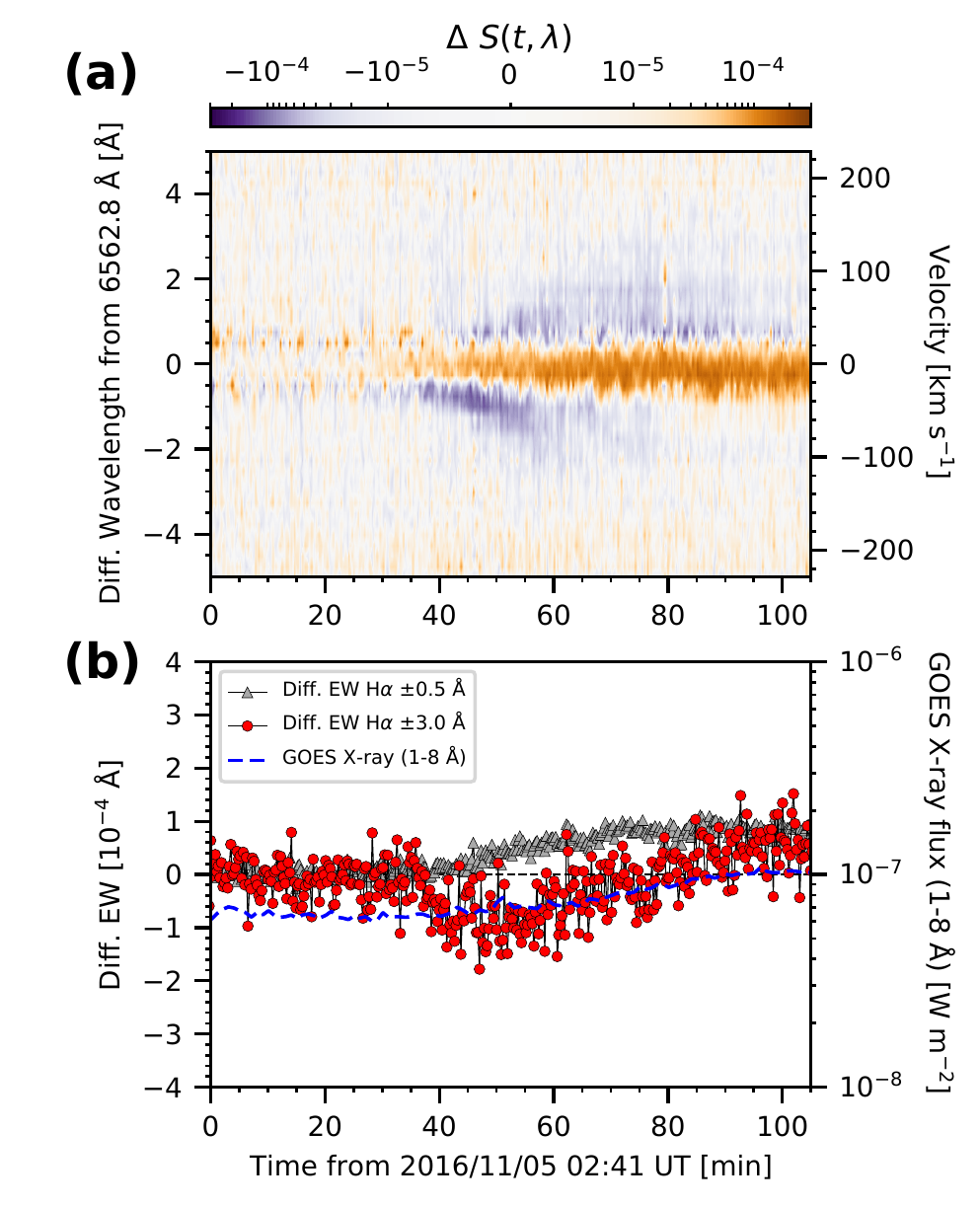}
\caption{The same as Figure \ref{fig:2017_0402_C}, but for the event (5), the filament eruption on 2016 November 5. In addition to the differenced equivalent widths of H$\alpha\pm3.0$~{\AA}, the differenced equivalent widths of H$\alpha\pm0.5$~{\AA} are also plotted as gray triangles in panel (b). 
\label{fig:2016_1105}}


\end{figure}

\begin{figure}
\centering
\includegraphics[width=8cm]
{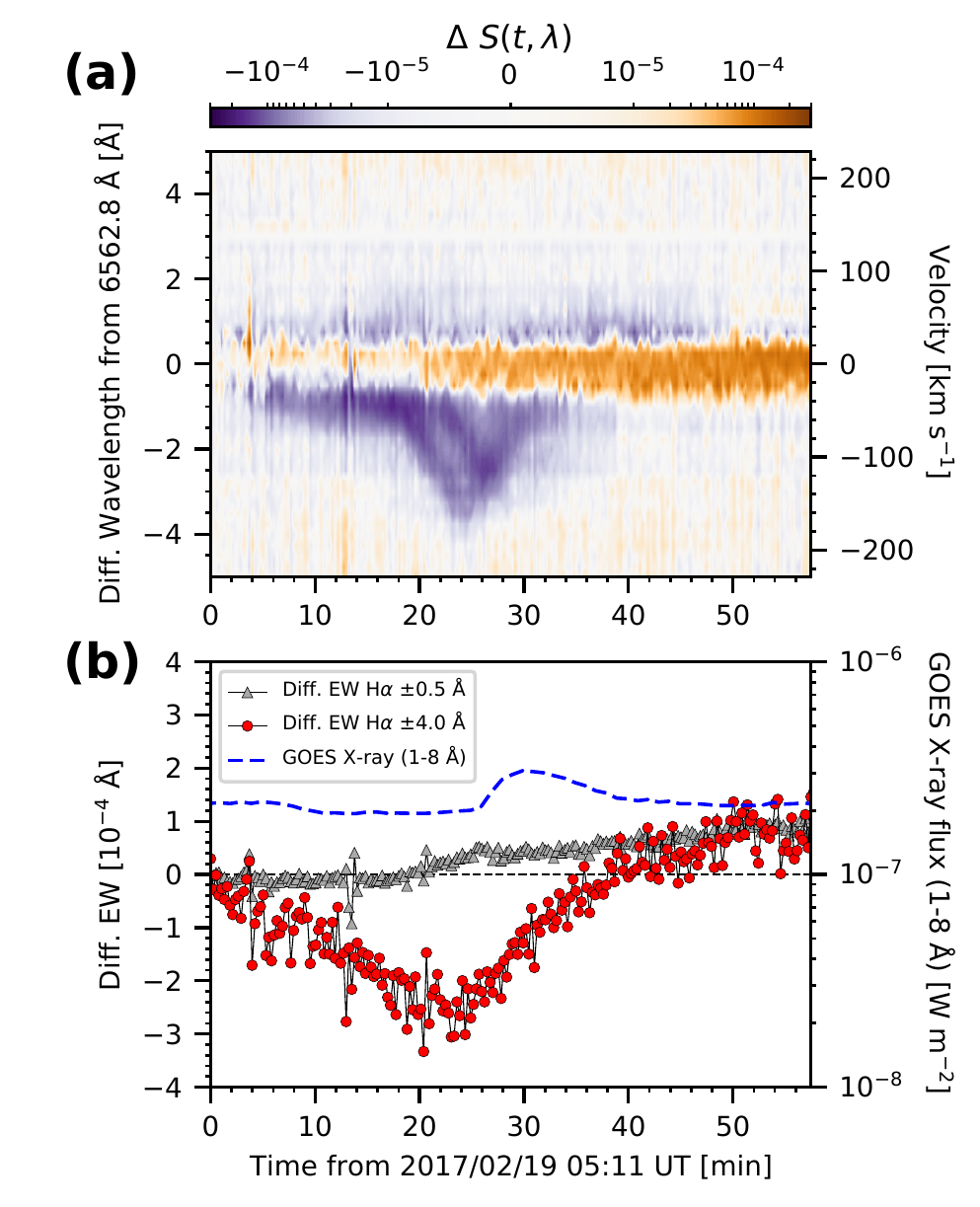}
\caption{The same as Figure \ref{fig:2017_0402_C}, but for event (6), the filament eruption and the two-ribbon flare on 2017 February 19. 
In addition to the differenced equivalent widths of H$\alpha\pm3.0$~{\AA}, the differenced equivalent widths of H$\alpha\pm0.5$~{\AA} are also plotted as gray triangles in panel (b), as in Figure \ref{fig:2016_1105} (b). 
\label{fig:2017_0219}}
\end{figure}

\begin{figure}
\centering
\includegraphics[width=8cm]
{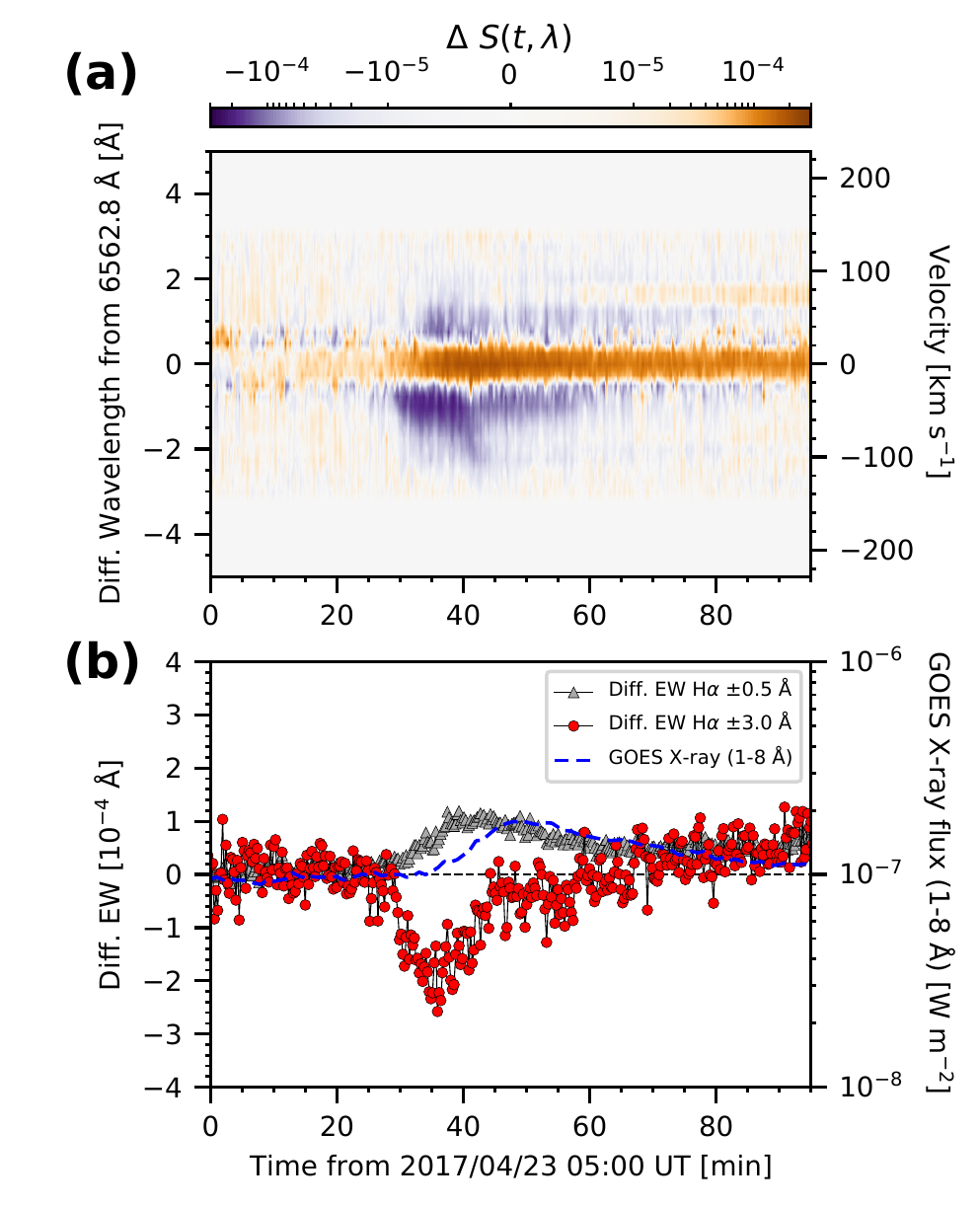}
\caption{The same as Figure \ref{fig:2017_0402_C}, but for event (7), the filament eruption and the two-ribbon flare on 2017 April 23. 
In addition to the differenced equivalent widths of H$\alpha\pm3.0$~{\AA}, the differenced equivalent widths of H$\alpha\pm0.5$~{\AA} are also plotted as gray triangles in panel (b), as in Figure \ref{fig:2016_1105} (b) and Figure \ref{fig:2017_0219} (b).
We note that only H$\alpha\pm3.0$~{\AA} data were available for performing the Sun-as-a-star analysis in this event.
\label{fig:2017_0423}}


\end{figure}

\begin{figure}
\centering
\includegraphics[width=8cm]
{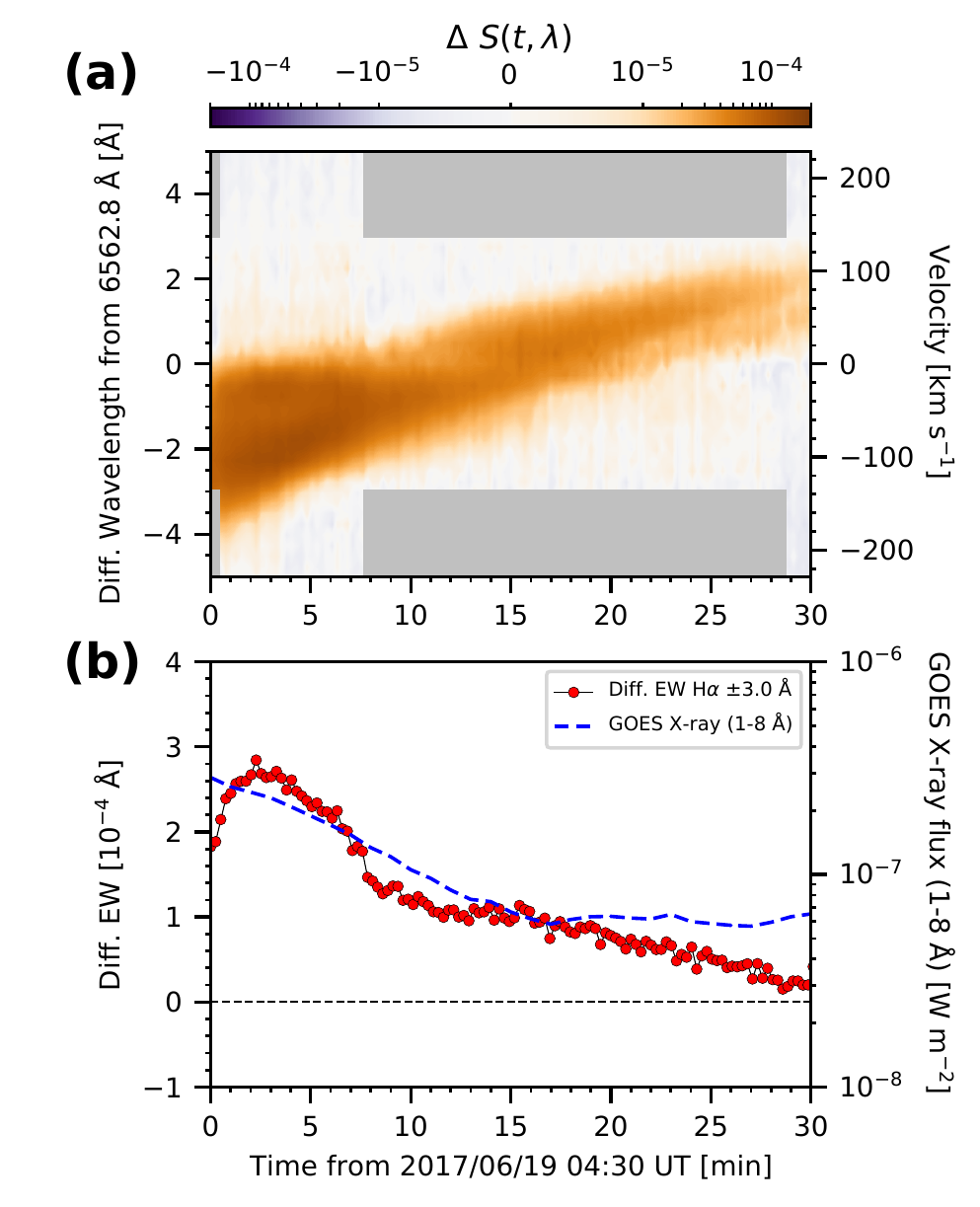}
\caption{The same as Figure \ref{fig:2017_0402_C}, but for event (8), the prominence eruption on 2017 June 19. We note that the peak time of \textit{GOES} soft X-ray (1$-$8 {\AA}) is before 2017 June 19 04:30 UT.
\label{fig:2017_0619}}

\end{figure}

\begin{figure}

\centering
\includegraphics[width=8cm]
{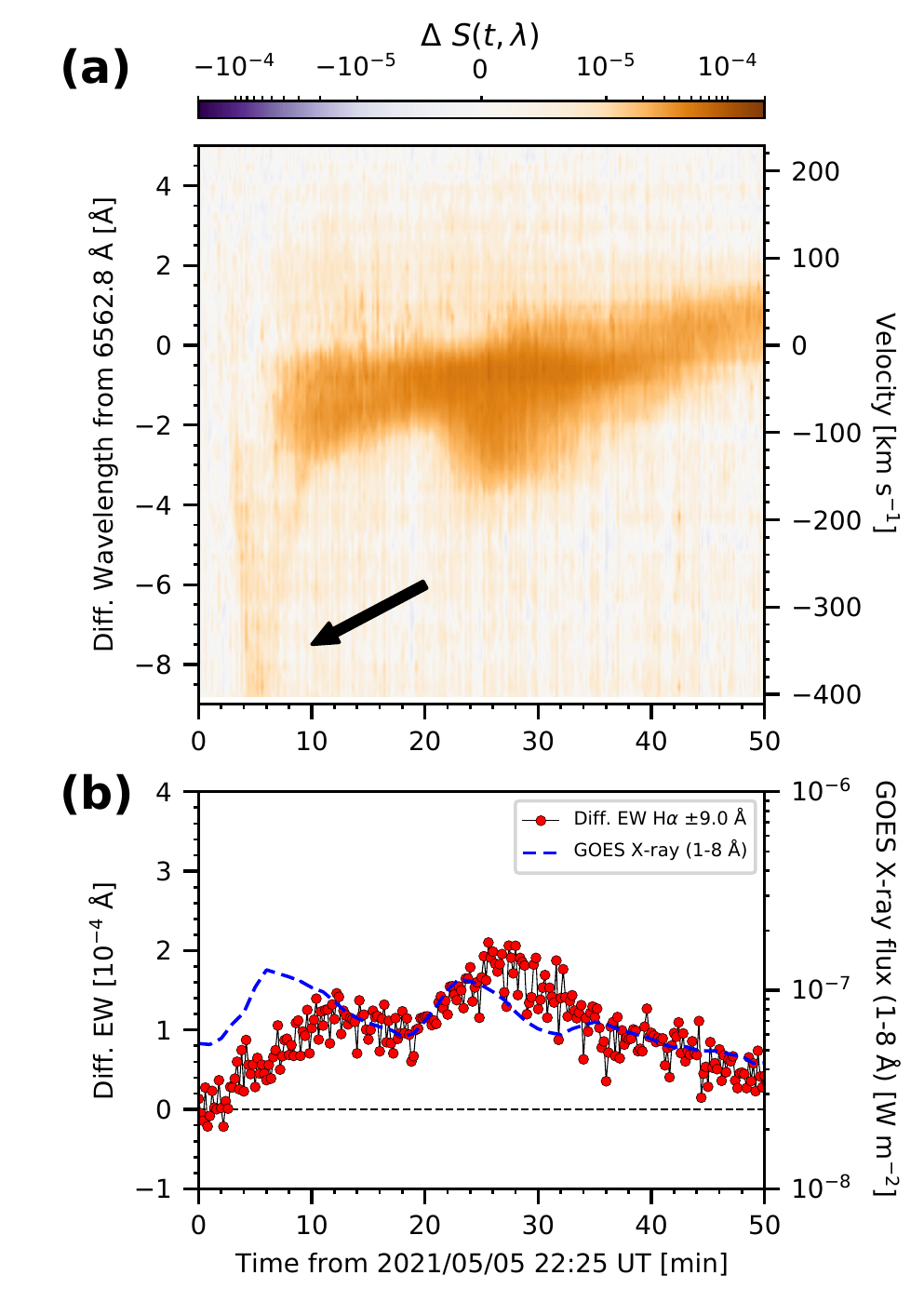}
\caption{The same as Figure \ref{fig:2017_0402_C}, but for event (9), the prominence eruption on 2021 May 5. In order to cover the fast component of the first eruption, we have extended the range of the blue wing to $-9.0$ {\AA} in panel (a). The black arrow in panel (a) indicates a fast component of this eruption.
\label{fig:2021_0505}}
\end{figure}

We also compared the time evolution of H$\alpha$ differenced equivalent widths and \textit{GOES} soft X-ray data.
In event (1)-(4) [flares], light curves of H$\alpha$ and \textit{GOES} soft X-ray are similar in the time scales of the impulsive phases and the peak times. This similarity is consistent with the standard flare model in which both H$\alpha$ and soft X-ray emissions are related to
chromospheric bombardment. On the other hand, in decay phases H$\alpha$ light curves sometimes show absorption due to line-of-sight motions of plasmas as previously reported by \citet{NamekataETAL2022a} and differ from soft X-ray light curves [especially in Figures \ref{fig:2017_0402_C} (b) and \ref{fig:2021_0420} (b)]. For the same reason, in event (5)-(7) [filament eruptions], H$\alpha$ light curves including both emission and absorption differs from soft X-ray light curves [red circles in Figures \ref{fig:2016_1105} (b), \ref{fig:2017_0219} (b), and \ref{fig:2017_0423} (b)]. In event (8) and (9) [prominence eruption], the peak time of H$\alpha$ delayed that of soft X-ray [Figures \ref{fig:2017_0619} (b) and \ref{fig:2021_0505} (b)]. The flare footpoints were occulted in these two events and H$\alpha$ emissions are purely caused by eruptions of off limb prominences. It takes much time for the prominence to expand to occupy a large area so the H$\alpha$ peak lagged behind the soft X-ray peak. This time lag may be the key to identifying whether H$\alpha$ emission comes from a flare or a prominence eruption but more statistics are needed to confirm its universality.

We highlight another important result: the cancellation between the emission due to an M5.3-class flare that occurred on the solar limb and the absorption due to the filament eruption. Out of the four flare events ($\geq$ C8.0), events (1), (2), and (3) were associated with H$\alpha$ filament eruptions. In the dynamic spectrum of event (1), which occurred near the disk center, clear absorption signals appears that corresponded to a surge. In contrast, the dynamic spectra of event (2) and event (3), which occurred around the limb, do not show clear absorptions. 
For event (2), the absorption caused by the eruption on the solar disk was cancelled or decreased by the flare emission near the H$\alpha$ line center because of the small LOS velocity of the filament.
On the other hand, some stellar flares show drastically broad emissions in their H$\alpha$ spectra (e.g., \citealt{NamekataETAL2020b} reported that a superflare on the M dwarf AD Leonis exhibited an H$\alpha$ line width up to 14 {\AA}.). 
Such broad emissions can cancel or decrease absorptions from filament eruptions with high LOS velocities. 
Thus, we deduce that for some stellar (super-) flares with broad H$\alpha$ emissions, 
filament eruptions on the stellar disks may not be detected in the H$\alpha$ spectra, even if filaments have high LOS velocities. 

As mentioned in summary (i), relatively prominent flares ($\geq$ C8.0) showed red asymmetry as signatures of chromospheric condensation in their impulsive phases. Such red asymmetry has been observed in other chromospheric lines. For example, \citet{GrahamETAL2020} analyzed the impulsive phase of a X-class solar flare and they reported that multiple chromospheric lines of Fe I, Fe II, Mg II, C I, and Si II  all showed red asymmetry. They also performed a radiative hydrodymamical simulation and showed that the red asymmetry can be explained by the collision of the electron beam with the solar chromosphere.
In addition, \citet{TianChen2018} reported that transition-region line of Si IV showed red asymmetry persisting in both impulsive and decay phases of a M-class solar flare. As for the stellar case, \citet{WoodgateETAL1992} reported that a stellar flare on the red dwarf star AU Mic showed red asymmetry in Ly$\alpha$ during the impulsive phase. This stellar observation suggests that chromospheric condensation also occurs during the impulsive phases of stellar flares. 

\citet{SekiETAL2019b} provided the catalogue of filament disappearance observed by SMART/SDDI, and they reported that the three filament eruptions in event (4), event (5) and event (6) evolved to become CMEs.  
On the other hand, the present study showed that red-shifted absorptions, which come from falling plasmas,
appear in the Sun-as-a-star H$\alpha$ spectra of these three events.
This suggests that even if falling plasmas are detected as red-shifted absorptions in the observations of stellar filament eruptions, such filament eruptions may also develop into CMEs.
We need further investigation to clarify whether we can deduce the occurrence of CMEs from spatially integrated H$\alpha$ spectra. 

In recent stellar observations, some H$\alpha$ spectra of stellar flares on M/K-dwarfs exhibited emission with blue asymmetry \citep[e.g.,][]{VidaETAL2016,VidaETAL2019, MuhekiETAL2020,MaeharaETAL2021}. 
They proposed that a prominence eruption caused the blue asymmetry in their study. 
They considered that emission along the H$\alpha$ line center due to the superflare and blue-shifted emission due to the prominence eruption moving toward Earth may overlap to produce such emissions with the blue asymmetry.
In the present study we found that solar prominences erupting toward Earth also resulted in blue shifted emissions in the Sun-as-a-star H$\alpha$ spectra. Moreover, the amplitudes of these blue-shifted emissions were comparable to those of solar flares (C8.0 $\leq$ flare class $\leq$ M8.1).
This suggests that if we observe flares with prominence eruptions, such prominence eruptions can make non-negligible contributions to the formation of the H$\alpha$ spectra.
Thus, our results for solar-prominence eruptions support the suggestion that the blue asymmetry may come from a prominence eruptions. In the future, Sun-as-a-star analyses are needed for a solar-prominence eruption with a footpoint flare in order to investigate whether prominence eruptions really can cause the blue asymmetry observed in the stellar cases.

In event (9), the emission with the high velocity up to $-$9.0 {\AA} for the first eruption is fainter than the emission with the small velocity, in the Sun-as-a-star H$\alpha$ dynamic spectrum [Figure \ref{fig:2021_0505} (a)]. 
We have considered that since the fast plasmas in this event occupied only a small spatial scale, their contributions to the emission in the spatially integrated spectra became much smaller than the contributions from the slow plasmas, which had large spatial scales.
For stellar observations, in which the number of photons is much smaller than that for the Sun, emission from faster but smaller blobs of plasma may become invisible in the spatially integrated observational data.
This suggests that even if high-velocity emission is not recognized in stellar H$\alpha$ spectra, faster plasma may still exist in stellar active phenomena. 

The relationship between flare enegy (e.g., X-ray fluence) and CME kinetic energy has been previously investigated \citep[e.g.,][]{YashiroGopalswamy2009}. On the other hand, the relationship between flare energy and kinetic energy of filament eruption is not clear. In the present paper, there is not enough number of data to discuss it, but their relation should be investigated by increasing the number of samples.

As shown in this study, the signals of active events generally survive in spatially integrated H$\alpha$ spectra.
This means that time-resolved H$\alpha$ spectroscopic observations are strong tools for distinguishing what the stellar active events look like. However, the signals of some active events in H$\alpha$ spectra may be suppressed by cancellation between emissions and absorptions, as occurred in event (2). 
This suggests that in addition to H$\alpha$ observations, the observations of other wavelengths are necessary to deduce the
stellar activities more accurately. 
Also, the comparison of H$\alpha$ line and another wavelength such as EUV is useful to investigate signals of CMEs in spatially integrated H$\alpha$ spectrum \citep{HarraETAL2016, XuETAL2022}. 
Multi-wavelength spectroscopy from the Sun-as-a-star viewpoint will be necessary to achieve a deeper understanding of various active events on stars.

\begin{acknowledgments}
The authors thank the anonymous referee for constructive comments that significantly improved the quality of this paper. We wish to thank Drs. K. Shibata and Y. Notsu for fruitful discussions.
We express our sincere gratitude to the staff of Hida Observatory for development and maintenance of the instrument and daily observation. 
We would like to acknowledge the data use from
GOES, SDO, and SOHO.
SDO is a mission for NASA’s Living With a Star program. 
This research was carried out by the joint research program of the Institute for Space-Earth Environmental Research, Nagoya University. 
This research is supported by JSPS KAKENHI grant numbers 21H01131 (K.I. and A.A.) and 18J20048, 21J00316 (K.N.). 
The authors would like to thank Enago for the English language review.

\end{acknowledgments}

\software{astropy \citep{Astropy}}

\newpage

\clearpage

\clearpage

\clearpage

\clearpage


\clearpage

\clearpage

\clearpage

\clearpage

\clearpage


\clearpage


\end{document}